\documentclass[3p,,preprint,,12pt]{elsarticle}
\usepackage{amsmath, amsfonts, mathrsfs, latexsym,amssymb}
\usepackage{graphicx}
\journal{European Journal of Mechanics - B/Fluids}

\usepackage{bm}
\usepackage{lineno,hyperref} 
\usepackage{epstopdf}
\usepackage[none]{hyphenat}
\usepackage{soul,xcolor}
\usepackage[normalem]{ulem}

 \usepackage{subfigure}
\newcommand{\kvs}[1]{\textcolor{black}{#1}}

\begin{document}

\begin{frontmatter}
\title{Viscoelastic Effects on the Hydrodynamics of an Active Compound Particle}
\author[a1,a2]{KVS Chaithanya}
\ead{chaithanyakvs@gmail.com}
\author[a3,a4]{Sumesh P. Thampi\corref{c1}}
\ead{sumesh@iitm.ac.in}\cortext[c1]{Corresponding author}
\address[a1]{ School of Life Sciences, University of Dundee, Dundee DD1 5EH, United Kingdom.}
\address[a2]{School of Science and Engineering, University of Dundee, Dundee DD1 4HN, United Kingdom}
\address[a3]{ Department of Chemical Engineering, Indian Institute of Technology Madras, Chennai, 600036, India}
\address[a4]{The Rudolf Peierls Centre for Theoretical Physics, Parks Road, Oxford OX1 3PU, United Kingdom}
\begin{abstract}

Understanding the hydrodynamics of microswimmers in viscoelastic fluids and confined environments is crucial for interpreting their behaviour in natural settings and designing synthetic microswimmers for practical applications like cargo transport. In this study, we explore the hydrodynamics of a concentric active compound particle - a model microswimmer (a squirmer) positioned at the centre of a viscoelastic fluid droplet (a model cargo) suspended in another viscoelastic medium. We consider the Oldroyd-B constitutive model to characterize the fluids and employ a perturbative approach in the Deborah number to analyze viscoelastic effects analytically, assuming a small Capillary number so that the droplet remains spherical and does not deform. We examine three cases: (i) a squirmer confined within a viscoelastic fluid droplet suspended in a Newtonian fluid, (ii) a squirmer confined within a Newtonian fluid droplet suspended in a viscoelastic fluid, and (iii) a squirmer confined within a viscoelastic fluid droplet suspended in another viscoelastic fluid. Our findings reveal that the swimming speeds of the squirmer and the droplet are determined by the complex interplay of viscoelasticity, the size ratio of the droplet to the squirmer (confinement strength), and the viscosity ratio of the surrounding fluid to the droplet fluid. A critical aspect of this interaction is the positioning of stagnation points within the fluid flow, which governs the distribution of polymeric stress. This distribution, in turn, plays a crucial role in determining the influence of viscoelasticity on the squirmer's dynamics. Our analysis suggests that viscoelastic effects can either enhance or hinder the swimming speed of the squirmer when confined in a droplet, depending on the specific configuration of the system, thus providing insights into the swimming behaviour of microswimmers in complex fluids and confinements.

\end{abstract}
\end{frontmatter}

\section{Introduction}\label{intro}

Many microorganisms navigate through complex fluids and confined environments, fundamental to their survival and reproduction. For instance, spermatozoa encounter a range of non-Newtonian fluid behaviours during their journey through the female reproductive tract towards the oocyte \cite{rutllant2005ultrastructural,miller2018epic}. The rheological properties of the fluid and the nature of confinement can significantly influence sperm motility, directly affecting the likelihood of successful fertilization \cite{lopez1994sperm,striggow2020sperm}. Similarly, pathogenic bacteria like \textit{Helicobacter pylori} move through the mucus lining of the stomach, adapting their motility strategies to persist and cause infection \cite{montecucco2001living,celli2009helicobacter,salama2013life}. Microorganisms such as algae and protozoa also encounter non-Newtonian fluids in various ecological niches, from muds to polysaccharide-rich plant tissues \cite{ferris1996dynamics,goldstein2015green}. Additionally, synthetic microswimmers designed for targeted drug delivery or bioremediation often face challenges navigating through complex fluids in confined spaces \cite{qiu2014swimming,wu2020medical,llacer2021biodegradable,murali2022advanced,mo2023challenges}. Therefore, understanding the dynamics of swimming microorganisms (microswimmers) in such environments is essential for gaining insights into their behaviour, optimizing their performance in practical applications, and providing a better understanding of infection mechanisms.\\

Microswimmers navigate within the realm of low Reynolds numbers, where viscous forces outweigh inertial forces \cite{childress1981mechanics,lauga2020fluid}. Consequently, unlike macroscopic organisms, microswimmers cannot utilize inertial forces to propel themselves; rather, they rely on viscous forces along with shape anisotropy to self-propel. Moreover, the hydrodynamics of Newtonian fluids (simple fluids), such as water, where the deviatoric stress is linearly related to strain rate, are governed by time-reversible equations of motion \cite{Camille2015book,leal2007advanced}. Achieving net displacement in such fluids requires adopting nonreciprocal kinematics to break time-reversal symmetry \cite{purcell1977Life}. This principle, known as the \textit{scallop theorem}, asserts that organisms dependent on reciprocal kinematics for locomotion cannot achieve net displacement in simple fluids at zero inertia. However, in non-Newtonian fluids, where the relationship between deviatoric stress and strain rate is non-linear, reciprocal motion can result in net displacement \cite{qiu2014swimming}.\\

Several studies have demonstrated that the nonlinear response of viscoelastic fluids significantly influences microswimmer dynamics. For instance, experiments with synthetic helical microswimmers, particularly those driven by external forces, have revealed that viscoelastic fluids can enhance propulsion by generating additional elastic stresses that interact with the microswimmer's structure \cite{dreyfus2005microscopic,wang2014acoustic,bechinger2016active,nishiguchi2018flagellar,quashie2023propulsion}. Furthermore, research on pathogenic bacteria such as \textit{Leptospira} \cite{kaiser1975enhanced,nakamura2022motility} and \textit{Borrelia burgdorferi} \cite{moriarty2008real,hyde2017borrelia,kimsey1990motility} - responsible for Weil’s disease and Lyme disease, respectively - has shown that these organisms swim more effectively in viscoelastic environments, aiding their movement through host tissues.  \kvs{On the other hand, the swimming speed of \textit{Caenorhabditis~elegans} in a viscoelastic fluid is found to be smaller compared to that in a Newtonian fluid, even when the shear viscosity is the same~\cite{shen2011undulatory}.}  These findings underscore the importance of viscoelasticity in both natural and synthetic systems, and reveal how complex fluid rheology can either facilitate or hinder microswimmer motion.\\

Complementing the experiments, theoretical studies have also provided insights into how viscoelasticity impacts the dynamics of microswimmers. Techniques such as the Lorentz reciprocal theorem \cite{ha1896eene,lorentz_trans,masoud2019reciprocal} and perturbation analyses using models like Giesekus or Oldroyd-B fluids, combined with the squirmer model, have been widely employed to investigate how weak viscoelastic effects influence the swimming speeds of different types of swimmers, including pushers, pullers, and neutral swimmers \cite{de2015locomotion,datt2019note,binagia2020swimming,housiadas2021squirmers}. In contrast, numerical simulations are often used to explore more pronounced viscoelastic effects \cite{zhu2012self,li2014effect,ouyang2023swimming,kobayashi2023direct}, particularly in scenarios where analytical methods are less effective. These approaches are crucial for guiding the design of synthetic microswimmers for various applications. However, both theoretical and numerical studies have predominantly focused on single-fluid systems.\\

An active compound particle, a microswimmer confined within a fluid droplet \cite{reigh2017swimming,reigh2017two,shaik2018locomotion,chaithanya2020deformation,chaithanya2023active}, presents a multiphase configuration that is of significance for two key reasons. First, the droplet, with its local fluid properties differing from the surrounding medium, enables the investigation of how these localized changes influence the swimmer's dynamics. This is particularly relevant for understanding the behaviour of pathogenic organisms that alter their local environment, and it also provides insights into how gradients in fluid properties can impact the mobility and efficiency of microswimmers. Second, the confinement within the droplet allows for the analysis of how spatial constraints affect microswimmer behaviour, which is crucial for understanding how these swimmers carry and transport cargo~\cite{li2025inhalable}.  Previous studies \cite{reigh2017swimming,reigh2017two,shaik2018locomotion} have explored the dynamics of active compound particles, under the assumption that both the droplet and the surrounding fluids are Newtonian. \kvs{\citet{reigh2017swimming} and \citet{reigh2017two}~have analytically demonstrated that the swimming speed varies non-monotonically with the size ratio of the droplet to the swimmer, approaching the unbounded swimmer speed in two extreme cases: (i) when the droplet is much larger than the swimmer, and (ii) when the droplet size is nearly equal to that of the swimmer. On the other hand, the swimming speed decreases monotonically with an increase in the viscosity ratio of the surrounding fluid to the droplet fluid. Moreover, \citet{reigh2017swimming} identified that the droplet consistently moved slower than a swimmer that uses purely tangential surface actuation, making the concentric configuration an instantaneous configuration. \citet{shaik2018locomotion}~extended these investigations by analyzing the dynamics of a swimmer within a surfactant-laden droplet, considering both concentric and eccentric configurations. They showed that surfactant redistribution on the droplet surface could either enhance or reduce the swimming speed, depending on the swimmer's eccentricity. Eccentric configurations introduced asymmetries that altered the balance of drag and thrust forces, and in some cases, led to the existence of fixed points at eccentric locations where the droplet speed matched the swimmer’s speed.} In this study, we theoretically examine the effects of viscoelasticity on the hydrodynamics of an active compound particle, \kvs{considering an instantaneous configuration,} where a squirmer is positioned at the centre of a viscoelastic fluid droplet suspended in another viscoelastic fluid medium. We use the Oldroyd-B model as the constitutive relation for both the droplet and the surrounding fluids, assuming that the Deborah number - the ratio of the fluid's relaxation time to the flow time scale - is small. Additionally, we assume that the capillary number, defined as the ratio of viscous forces to surface tension, is small, ensuring the droplet remains spherical and does not deform. \\


This paper is organized as follows. In Sec.~\ref{mathform}, we present the mathematical formulation, including the squirmer model, the governing equations for fluid flow, and the relevant boundary conditions. Section~\ref{sol_procedure} details the solution procedure, beginning with the zeroth-order solution for a Newtonian fluid and extending to the first non-Newtonian correction using a perturbation approach. The results are discussed in Sec.~\ref{results}, where we analyze three distinct cases: (i) a squirmer confined within a viscoelastic fluid droplet suspended in a Newtonian fluid (Sec.~\ref{results1}), (ii) a squirmer confined within a Newtonian fluid droplet suspended in a viscoelastic fluid (Sec.~\ref{results2}), and (iii) a squirmer confined within a viscoelastic fluid droplet suspended in another viscoelastic fluid (Sec.~\ref{results3}). Finally, Sec.~\ref{conclu} concludes the paper by summarizing the key findings and discussing their implications for understanding microswimmer dynamics in complex fluids.

\section{Mathematical Formulation}\label{mathform}

\begin{figure}[tbh]
  \centering
  \includegraphics[width=0.4\linewidth]{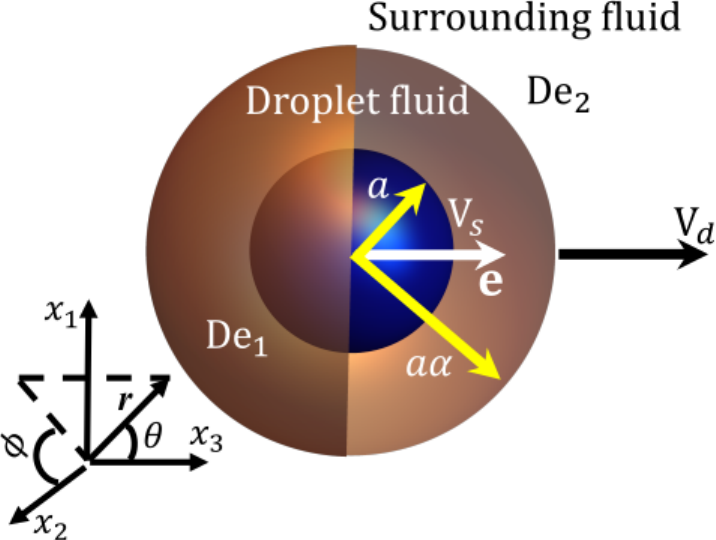}
  \caption{\label{fig:schematic} A schematic representation of an active compound particle - an active particle with radius $a$ encapsulated in a fluid droplet of radius $a\alpha$. The orientation of the active particle is denoted by $\mathbf{e}$, while De$_1$ and De$_2$ denote the Deborah number of the inner and outer fluids, respectively. The squirmer and the droplet swim in the $x_3$-direction, each with a distinct velocity: $V_s$ for the squirmer and $V_d$ for the droplet. Cartesian coordinates ($x_1, \ x_2, \ x_3$) and spherical coordinates ($r, \ \theta, \ \phi$) used in the analysis are shown.}
\end{figure}
In this study, we investigate the effect of viscoelasticity on the dynamics of a neutrally buoyant concentric active compound particle. This configuration consists of a microswimmer positioned at the centre of a viscoelastic droplet,  with the droplet suspended in another viscoelastic fluid as shown in Fig.~\ref{fig:schematic}.

\subsection{Squrimer - a model microswimmer}

\vspace{0.25cm}

The microswimmer is modelled using an axisymmetric spherical squirmer model initially proposed by Lighthill \cite{lighthill1952squirming} and Blake \cite{blake1971spherical}. This model describes the motion of the cilia on the microswimmer's surface as a slip velocity, which is a continuous function of the angle relative to the orientation vector, $\mathbf{e}$ (Fig.~\ref{fig:schematic}), expressed in terms of spherical harmonics as
\begin{equation}\label{eqn:squi_model}
    \mathbf{u}^{sq} = \sum_{n=1}^{n=\infty}  B_n V_n(\cos \theta) \mathbf{e}_\theta.
\end{equation}

Here, $V_n(\cos\theta) = \frac{-2}{n(n+1)}P_n^{1}(\cos\theta)$, where $P_n^{1}(\cos\theta)$ represents the associated Legendre Polynomials of the first kind, with $\theta$ being the polar angle measured relative to the axis of symmetry ($\mathbf{e}$). 
The coefficients $B_n$, known as squirming modes, govern the microswimmer's behaviour. Typically, only the first two squirming modes, namely $B_1$ and $B_2$, are considered in the studies. This choice stems from the behaviour observed in a Newtonian fluid, where $B_1$ and $B_2$ correspond to source and force dipoles, respectively, with flow associated with $B_1$ mode decaying as $1/r^3$ and that of $B_2$ mode decaying as $1/r^2$.  The higher-order modes ($n>3$) generate flows that decay as $1/r^{2n+1}$.  Consequently, only the first two modes significantly contribute to the far-field velocity produced by the squirmer, making them the primary focus of locomotion problems, including our present study. \\

Additionally, the ratio $\beta = B_2/B_1$ is often used to differentiate between different types of microswimmers. Specifically, when $\beta>0$, a squirmer mimics the behaviour of puller-type microswimmers like \textit{Chlamydomonas}, while $\beta<0$ mirrors the behaviour of pushers like \textit{E. coli}. In an unbounded Newtonian fluid, the swimming speed of a squirmer is given by $V_s^{\text{UN}} = \frac{2}{3} B_1$, indicating that the swimming speed is same for all types of swimmers regardless of the sign and value of $\beta$ (or $B_2$).
    
\subsection{Governing equations for fluid flow}

The characteristic length and velocity scales associated with microswimmers and compound particles are small, resulting in a typical Reynolds number, the ratio of inertial to viscous forces, of $\mathcal{O}(10^{-6})$. Therefore, in our analysis, we neglect fluid inertia and assume that the Stokes and continuity equations govern the velocity and pressure fields inside and outside the droplet, as given by,

\begin{equation}\label{eqn:d_gov_eq}
     \bm\nabla\cdot\bm{\tau^*}_{k} = \bm{\nabla} p^*_{k}, \quad \bm\nabla \cdot \mathbf{u}^*_{k} = 0,
\end{equation}

where $\mathbf{u}^*_{k}$ represents the velocity field and $p^*_{k}$ denotes the dynamic pressure in each fluid, with $k = 1, 2$ representing the droplet and surrounding fluids, respectively. In addition, $\bm{\tau}^*_k$ is the deviatoric stress tensor in each fluid, determined by a constitutive equation. \\

In this work, we consider the Oldroyd-B model as the constitutive relation for viscoelastic fluids. The corresponding deviatoric stress tensor for the droplet and surrounding fluids is given by,

\begin{equation}
    \bm{\tau}^*_{k}+\zeta_{k} \stackrel{\nabla}{\bm{\tau}^*_{k}}= 2\eta_{k}\Big(\mathbf{D}^*_{k}+\zeta_{k}(1-m_{k})\stackrel{\nabla}{\mathbf{D}}{^*_{k}}\Big).
\end{equation}

Here, $\zeta_{k}$ is the relaxation time of the fluid, $\eta_k = \eta_{s,k}+\eta_{p,k}$ is the zero-shear fluid viscosity which is the sum of solvent viscosity ($\eta_{s,k}$) and polymer viscosity ($\eta_{p,k}$), $m_k = \eta_{p,k}/\eta_{k}$ is the polymer viscosity ratio, and $\mathbf{D}_k^*=\frac{1}{2}\Big(\mathbf{\nabla^* u}_k^*+(\mathbf{\nabla^* u}_k^*)^T\Big)$ represents the rate of strain tensor, where the superscript $T$ denotes the transpose. $\stackrel{\nabla}{\mathbf{D}}{^*_{k}}$ is the upper-convected derivative of the velocity-gradient tensor $\mathbf{D}^*_k$, expressed as,
\begin{equation}
    \stackrel{\nabla}{\mathbf{D}}{^*_{k}} = \frac{\partial \mathbf{D}_k^*}{\partial t^*}+\mathbf{u}_k^* \cdot \bm{\nabla}^*\mathbf{D}_k^*-\mathbf{D}_k^*\cdot\bm{\nabla}^*\mathbf{u}_k^*-(\bm{\nabla}^*\mathbf{u}_k^*)^T\cdot\mathbf{D}_k^*.
\end{equation}
Equation~\ref{eqn:d_gov_eq} can be transformed into a dimensionless form using the size of a squirmer, $a$, as a characteristic length scale, $B_1$ as the characteristic velocity scale, and $\eta_{2}B_1/a$ as the characteristic stress scale. The non-dimensional Stokes and continuity equations obtained are given by,
\begin{equation}\label{eqn:nd_gov_eq}
     \bm\nabla\cdot\bm{\tau}_{k} = \bm{\nabla} p_{k}, \quad \bm\nabla \cdot \mathbf{u}_{k} = 0,
\end{equation}
and the corresponding deviatoric stress tensor is given by,
\begin{equation}\label{eqn:nd_const_eq}
  \begin{aligned}
    \bm{\tau}_{1}+\text{De}_{1}\stackrel{\nabla}{\bm{\tau}}_{1} &= (2/\lambda)(\mathbf{D}_{1}+\text{De}_{1}(1-m_{1})\stackrel{\nabla}{\mathbf{D}}_{1}), \\
    \bm{\tau}_{2}+\text{De}_{2} \stackrel{\nabla}{\bm{\tau}}_{2} &= 2(\mathbf{D}_{2}+\text{De}_{2}(1-m_{2})\stackrel{\nabla}{\mathbf{D}}_{2}), 
    \end{aligned}
\end{equation}  
where $\lambda = \eta_2/\eta_1$ is the viscosity ratio, and De$_k = \zeta_k B_1/a$ is the Deborah number which is the ratio between the relaxation time of the fluid ($\zeta_k$) and the flow time scale ($a/B_1$). Additionally, De$_2=$ De$_1$/$\gamma$, where $\gamma = \zeta_1/\zeta_2$ is referred to as the elasticity ratio.
\subsubsection{Boundary conditions}
\label{sec:gen_BC}

\begin{enumerate}[i.]
    \item In the laboratory frame, the fluid velocity in the far field decays as
        \begin{equation} \label{eqn:gen_far_field_bc}
            \mathbf{u}_2(r\to\infty) = \mathbf{0}.
        \end{equation}
    \item The no-slip boundary condition on the squirmer surface is given by,
        \begin{equation}
            \mathbf{u}_1 = \mathbf{V}_s + \mathbf{u}^{sq},
        \end{equation}
        where $\mathbf{V}_s$ is the unknown swimming speed of the microswimmer, and $\mathbf{u}^{sq} = \sin\theta+\beta\cos\theta\sin\theta$ is the surface slip velocity (Eqn.~\ref{eqn:squi_model}).
    \item On the droplet surface, the continuity of both velocity and traction is maintained. Here, we assume that the  Capillary number, Ca$ = \lambda \mu B_1/\sigma$ is small, where $\sigma$ is the interfacial tension. This ensures that the interfacial tension is sufficiently large to maintain the spherical shape of the droplet. 

    Therefore, the boundary conditions on the droplet surface ($r = \alpha$) are given by,

    \begin{equation}
    \begin{aligned}        
        &u_{1,r} = u_{2,r} = V_d \cos \theta, \\ 
        &u_{1,\theta} = u_{2,\theta}, \text{ and} \\
        &\mathbf{n} \cdot \bm{\tau}_1 \cdot \mathbf{t} = \mathbf{n} \cdot \bm{\tau}_2 \cdot \mathbf{t}.     
    \end{aligned}        
    \end{equation}
    Here $u_{k,r}$ and $u_{k,\theta}$ are the radial and tangential components of the fluid velocity, respectively,  $V_d$ is the unknown swimming speed of the droplet, $\mathbf{n}$ and $\mathbf{t}$ are the unit normal and tangent vectors to the interface.
    \item Finally, the net hydrodynamic forces exerted on both the swimmer ($\mathbf{F}_s$) and the droplet interface ($\mathbf{F}_d$) are zero,
    \begin{equation}\label{eqn:force_free_gen}
    \begin{aligned}  
        \mathbf{F}_{s} &=\int_{4\pi} \mathbf{T}_{1} \cdot \mathbf{n} \ \text{d}S = 0, \\
        \mathbf{F}_{d} &= \int_{4\pi\alpha^2} \mathbf{T}_{2} \cdot \mathbf{n} \ \text{d}S = 0,
    \end{aligned}        
    \end{equation}
 
    where $\mathbf{T}_k = -p_k \mathbf{I} + \bm{\tau}_k$ is the total stress. Equations~\ref{eqn:force_free_gen} are used to determine the swimming speeds of the squirmer ($V_s$) and the droplet ($V_d$) in the following calculations.
\end{enumerate}

\subsection{Solution procedure}
\label{sol_procedure}

Equations~\ref{eqn:nd_gov_eq}, along with the constitutive equations (Eqn.~\ref{eqn:nd_const_eq}) are nonlinear and lack a general solution for arbitrary values of Deborah number (De$_k$) given the boundary conditions, Eqns.~\ref{eqn:gen_far_field_bc}--\ref{eqn:force_free_gen}. However, using a regular perturbation approach, it is possible to linearize the equations and solve them under the assumption of a small Deborah number, De$_k$ $\ll 1$. This assumption implies that the relaxation time of the fluid is much shorter than the characteristic time for the fluid flow, i.e., $\zeta_k \ll a/B_1$. Here, using De$_2$ as a perturbation parameter, we expand any dependent variable $\xi$ in the following form:
\begin{equation}
    \xi = \xi(0)+ \text{De}_2 \xi(1)+\dotsb,
\end{equation}
where $\xi(0) = \xi^{N}$ denotes the zeroth-order term which corresponds to a Newtonian fluid, and $\xi(1) = \xi^{\text{De}_2}$ represents the first non-Newtonian correction due to viscoelasticity. Equations corresponding to different orders in De$_2$ are obtained by substituting the asymptotic expansion for all dependent variables into the governing equations and boundary conditions. Moreover, we assume that $\text{De}_1$ and  $\text{De}_2$ are of the same order of magnitude, implying $\gamma \sim 1$.

\subsubsection{Zeroth-order solution - Newtonian fluid} \label{sec:zeroeth_order}

The zeroth-order solution corresponds to that of Newtonian fluid and has been discussed in detail by \citet{reigh2017swimming}, and here we give the details in brief. The governing equations for the fluid flow at zeroth-order are given by,

\begin{equation}\label{eqn:zeroth_order_ge}
\begin{aligned}
    \bm{\nabla} p_1^{\text{N}} &= (1/\lambda) \bm{\nabla}^2\mathbf{u}_1^{\text{N}}, \quad \bm{\nabla} \cdot \mathbf{u}_1^{\text{N}} = 0, \\
    \bm{\nabla} p_2^{\text{N}} &= \bm{\nabla}^2\mathbf{u}_2^{\text{N}}, \quad \bm{\nabla} \cdot \mathbf{u}_2^{\text{N}} = 0.
\end{aligned}
\end{equation}


Given the axisymmetric nature of the problem, we take the curl of the momentum equations to eliminate the pressure terms and introduce a stream function \(\Psi_k\), which simplifies the system into a single scalar equation for \(\Psi_k\) as follows:
\begin{align}\label{eqn:zeroth_order_str}
\begin{split}
    E^2(E^2\Psi_1^{\text{N}}) &= 0, \\
    E^2(E^2\Psi_2^{\text{N}}) &= 0,
\end{split}
\end{align}

where the linear operator $E^2$ is given by $E^2 = \frac{\partial^2}{\partial r^2}+\frac{(1-\mu^2)}{r^2}\frac{\partial^2}{\partial \mu^2}$ with $\mu = \cos \theta$. The radial and tangential components of the velocity field are related to the stream function via,
\begin{equation}\label{eqn:vel_str_relations}
    u_{k,r} = -\frac{1}{r^2} \frac{\partial \Psi_{k}}{\partial \mu}, \quad u_{k,\theta} = -\frac{1}{r\sqrt{1-\mu^2}} \frac{\partial \Psi_k}{\partial r}.
\end{equation}

Equations~\ref{eqn:zeroth_order_str} are solved using the following zeroth-order boundary conditions,
\begin{align}
\begin{split}
    &\mathbf{u}_2^{\text{N}} = \mathbf{0} \quad \text{at} \ r \to \infty, \\      
    &u_{1,r}^{\text{N}} = u_{2,r}^{\text{N}} = V_d^{\text{N}} \cos \theta \quad \text{at } r = \alpha, \\
    &u_{1,\theta}^{\text{N}} = u_{2,\theta}^{\text{N}}  \quad \text{at } r = \alpha, \\
    &\mathbf{n} \cdot \bm{\tau}_1^{\text{N}} \cdot \mathbf{t} = \mathbf{n} \cdot \bm{\tau}_2^{\text{N}} \cdot \mathbf{t}\quad \text{at } r = \alpha, \\ 
    &u_{1,r}^{\text{N}} = V_s^{\text{N}} \cos \theta \quad \text{at } r = 1 \text{ and} \\   
    &u_{1,\theta}^{\text{N}} = (1-V_s^{\text{N}}) \sin \theta +  \beta \cos \theta \sin \theta  \quad \text{at } r = 1. \\
    \end{split}
\end{align}

The swimming speeds of the squirmer and the droplet at the zeroth order are determined via,
\begin{align}
\begin{split}    
    \mathbf{F}_{s}^N &=\int_{4\pi} \mathbf{T}_{1}^N \cdot \mathbf{n} \ \text{d}S = 0, \\
    \mathbf{F}_{d}^N &= \int_{4\pi\alpha^2} \mathbf{T}_{2}^N \cdot \mathbf{n} \ \text{d}S = 0.
\end{split}
\end{align}

The expressions for the leading-order stream functions under these boundary conditions are as follows:

\begin{align}
    \Psi_1^{\text N} &= \sum_{l=1}^{2}\left[A_{l1}^{\text N} r^{l+3}+B_{l1}^{\text N} r^{l+1}+C_{l1}^{\text N} r^{2-l} +D_{l1}^{\text N} r^{-l}\right] \mathcal{Q}_l(\mu),\label{eqn:zeroth_str_sol_1} \\
    \Psi_2^{\text N} &= \sum_{l=1}^{2}\left[C_{l2}^{\text N} r^{2-l} +D_{l2}^{\text N} r^{-l}\right] \mathcal{Q}_l(\mu), \label{eqn:zeroth_str_sol_2}
\end{align}

where \(\mathcal{Q}_l(\mu) = \int_{-1}^\mu P_l(\mu) \, d\mu\), and \(P_l(\mu)\) represents the Legendre polynomial. The terms $A_{lk}^{\text N}, \ B_{lk}^{\text N}, \ C_{lk}^{\text N}, \ D_{lk}^{\text N}$ are constants. The equations determining these constants, derived by applying the boundary conditions, are provided in the~\ref{sec:AppendixA}.\\

The swimming speed of the squirmer and the droplet at the zeroth order is given by \cite{reigh2017swimming},

\begin{equation}
    V_s^\text{N} = \frac{\lambda(2\alpha^5-5\alpha^2+3)+3\alpha^5+5\alpha^2-3}{2(\alpha^5-1)\lambda+3\alpha^5+2}.
\end{equation}

\begin{equation}
    V_d^\text{N} = \frac{5\alpha^2}{2(\alpha^5-1)\lambda+3\alpha^5+2}.
\end{equation}

Similar to the case of an unbounded squirmer, the swimming speeds are independent of $\beta$ \cite{lighthill1952squirming,blake1971spherical} here. However, unlike an unbounded squirmer, they depend on the size and viscosity ratio of the compound particle. As noted by \citet{reigh2017swimming}, the swimming speeds of the microswimmer varies non-monotonically with the size ratio, approaching the unbounded squirmer's speed in a Newtonian fluid, $V_s^N$, as $\alpha$ approaches $1$ or $\infty$ for $\lambda \neq 1$. The limit $\alpha \to 1$ corresponds to a squirmer in an unbounded fluid with viscosity $\eta_2$, and the limit $\alpha \to \infty$ corresponds to the squirmer in an unbounded fluid with a viscosity of $\eta_1$. When $\lambda = 1$ there is no correction to the swimming speed due to the droplet confinement. Moreover, the swimming speed decreases with an increase in the viscosity of the suspending fluid or the viscosity ratio. Consequently, the presence of a confining droplet can either increase or decrease the swimming speed of the squirmer, depending on the viscosity ratio. On the other hand, the droplet consistently swims slower than the squirmer, and its speed monotonically decreases with an increase in either the size ratio or the viscosity ratio.

\subsubsection{$\mathcal{O}(\text{De}_2)$ solution - first non-Newtonian correction}

\label{sec:O_De_soln}

In this section, we present the governing equations and boundary conditions at $\mathcal{O}(\text{De}_2)$ and derive the corresponding corrections to the flow field and swimming speeds. \\

The governing equations for the fluid flow at $\mathcal{O}(\text{De}_2)$ are given by,

\begin{align}\label{eqn:first_order_ge}
 \begin{split}
    \bm{\nabla} p_1^{\text De_2} &= (1/\lambda) \bm{\nabla}^2\mathbf{u}_1^{\text De_2}+\bm \nabla \cdot \bm{\tau}_{1p}^{\text De_2}, \quad \bm{\nabla} \cdot \mathbf{u}_1^{\text De_2} = 0, \\
    \bm{\nabla} p_2^{\text De_2} &= \bm{\nabla}^2\mathbf{u}_2^{\text De_2}+\bm \nabla \cdot \bm{\tau}_{2p}^{\text De_2}, \quad \bm{\nabla} \cdot \mathbf{u}_2^{\text De_2} = 0.
 \end{split}
\end{align}
Here, $\bm{\tau}_{kp}^{\text De_2}$ represents the polymeric stress due to the fluid viscoelasticity and are solely determined by the zeroth-order solution (Sec.~\ref{sec:zeroeth_order}),

\begin{equation}\label{eqn:first_order_poly_st}
    \begin{aligned}
    \bm{\tau}_{1p}^{\text De_2} &= -(2m_1\gamma/\lambda)\Big[\mathbf{u}_1\cdot\bm\nabla\mathbf{D}_1^{\text N}-(\bm \nabla \mathbf{u}_1^{\text N})^T\cdot\mathbf{D}_1^{\text N}-\mathbf{D}_1^{\text N}\cdot\bm\nabla\mathbf{u}_1^{\text N}\Big], \\
    \bm{\tau}_{2p}^{\text De_2} &= -2m_2\Big[\mathbf{u}_2^{\text N}\cdot\bm\nabla\mathbf{D}_2^{\text N}-(\bm \nabla \mathbf{u}_2^{\text N})^T\cdot\mathbf{D}_2^{\text N}-\mathbf{D}_2^{\text N}\cdot\bm\nabla\mathbf{u}_2^{\text N}\Big].
    \end{aligned}
\end{equation}

Similar to zeroth-order problem, we recast Eqns.~\ref{eqn:first_order_ge} in terms of stream function \cite{das2017effect} as,

\begin{equation}\label{eqn:first_order_str}
    \begin{aligned}
    E^2(E^2\Psi_1^{\text{De}_2}) &= \lambda r\sqrt{1-\mu^2}\Bigg[\bm \nabla \times (\bm \nabla \cdot \bm\tau_{1p}^{\text De_2})\Bigg]\cdot\mathbf{e}_\phi, \\
    E^2(E^2\Psi_2^{\text{De}_2}) & = r\sqrt{1-\mu^2}\Bigg[\bm \nabla \times (\bm \nabla \cdot \bm\tau_{2p}^{\text De_2})\Bigg]\cdot\mathbf{e}_\phi.
    \end{aligned}
\end{equation}
Here, $\mathbf{e}_\phi$ denotes the azimuthal unit vector.
Equations~\ref{eqn:first_order_str} are solved using the following first-order boundary conditions,

\begin{equation}\label{eqn:first_order_bc}
\begin{aligned}
    &\mathbf{u}_2^{\text{De}_2} = \mathbf{0} \quad \text{at} \ r \to \infty, \\      
    &u_{1,r}^{\text{De}_2} = u_{2,r}^{\text{De}_2} = V_d^{\text{De}_2} \cos \theta \quad \text{at } \ r = \alpha, \\
    &u_{1,\theta}^{\text{De}_2} = u_{2,\theta}^{\text{De}_2}  \quad \text{at } \ r = \alpha, \\  
    &\mathbf{n}\cdot\bm{\tau}_1^{\text{De}_2} \cdot \mathbf{t} = \mathbf{n}\cdot\bm{\tau}_2^{\text{De}_e} \cdot \mathbf{t}\quad \text{at } \ r = \alpha, \\ 
    &u_{1,r}^{\text{De}_2} = V_s^{\text{De}_2} \cos \theta \quad \text{at } \ r = 1 \text{ and} \\   
    &u_{1,\theta}^{\text{De}_2} = -V_s^{\text{De}_2} \sin \theta \quad \text{at } \ r = 1, \\
\end{aligned}
\end{equation}
where $\bm{\tau}_1^{\text{De}_2} = \frac{2}{\lambda} \mathbf{D}_1^{\text{De}_2} + \bm{\tau}_{1p}^{\text De_2}$, and $\bm{\tau}_2^{\text{De}_2} = 2\mathbf{D}_1^{\text{De}_2} + \bm{\tau}_{2p}^{\text De_2}$.\\

Additionally, the $\mathcal{O}(\text De_2)$ correction to the swimming speeds of the squirmer and the droplet are determined via,

\begin{equation}\label{eqn:first_order_force_free}
\begin{aligned}
    \mathbf{F}_{s}^{\text De_2}&=\int_{4\pi} \mathbf{T}_{1}^{\text De_2} \cdot \mathbf{n} \ \text{d}S = 0, \\
    \mathbf{F}_{d}^{\text De_2} &= \int_{4\pi\alpha^2} \mathbf{T}_{2}^{\text De_2} \cdot \mathbf{n} \ \text{d}S = 0,
\end{aligned}
\end{equation}
where $\mathbf{T}_k^{De_2} = -2p_k^{\text{De}_e}\mathbf{I}+ \bm{\tau}_k^{\text{De}_2}$.\\

Equations~\ref{eqn:first_order_str} can be further simplified using the zeroth order velocity field and Eqn.~\ref{eqn:first_order_poly_st} as,

\begin{equation}\label{eqn:first_order_str_simp}
    \begin{aligned}
    E^2(E^2\Psi_1^{\text{De}_2}) &= \sum_{i = 1}^{4} g_i(r)\mathcal{Q}_i(\mu), \\
    E^2(E^2\Psi_2^{\text{De}_2}) & =\sum_{i = 1}^{4} h_i(r)\mathcal{Q}_i(\mu), \\
    \end{aligned}
\end{equation}
where the coefficients $g_i(r)$ and $h_i(r)$ are given in the~\ref{sec:AppendixB}.\\

The $\mathcal{O}(\text{De}_2)$ solution to the stream functions obtained by solving Eqn.~\ref{eqn:first_order_str_simp} and satisfying the boundary conditions (Eqn.~\ref{eqn:first_order_bc}) is given by:
\begin{equation}\label{eqn:first_order_str_sol}
    \begin{aligned}
    \Psi_1^{\text{De}_2} &= \sum_{l=1}^{4}\left[A_{l1}^{\text{De}_2} r^{l+3}+B_{l1}^{\text{De}_2} r^{l+1}+C_{l1}^{\text{De}_2} r^{2-l} +D_{l1}^{\text{De}_2} r^{-l}+\mathcal{I}_{l1}^{\text{De}_2}(r)\right] \mathcal{Q}_l(\mu), \\
    \Psi_2^{\text{De}_2} &= \sum_{l=1}^{4}\left[C_{l2}^{\text{De}_2} r^{2-l} +D_{l2}^{\text{De}_2} r^{-l}+\mathcal{I}_{l2}^{\text{De}_2}(r)\right] \mathcal{Q}_l(\mu),
    \end{aligned}
\end{equation}

where $A_{lk}^{\text{De}_2}$, $B_{lk}^{\text{De}_2}$, $C_{lk}^{\text{De}_2}$, and $D_{lk}^{\text{De}_2}$ are constants. The equations that determine these constants, derived by applying the boundary conditions, are provided in the Supplementary Mathematica notebook. Additionally, the expressions for the coefficients $\mathcal{I}_{lk}^{\text{De}_k}$, which constitute the particular integral part of the solution to Eqns.~\ref{eqn:first_order_str_simp}, are provided in the~\ref{sec:AppendixC}.\\

Finally, the total swimming speeds of the squirmer and the droplet are given by:
\begin{equation}\label{eqn:}
    \begin{aligned}
    V_s &= V_s^{\text{N}} + \text{De}_2 V_s^{\text{De}_2}, \\
    V_d &= V_d^{\text{N}} + \text{De}_2 V_d^{\text{De}_2},
\end{aligned}
\end{equation}
where $V_s^{\text{De}_2}$ and $V_d^{\text{De}_2}$ represent the $\mathcal{O}(\text{De}_2)$ corrections to the swimming speeds of the squirmer and the droplet, respectively. These corrections are provided in the Supplementary Mathematica Notebook and are determined using Eqn.~\ref{eqn:first_order_force_free}.

\section{Results} \label{results}

To comprehensively understand the interplay between viscoelasticity and droplet-induced confinement on the dynamics of an active compound particle, we analyze three distinct cases: (i) a squirmer confined within a viscoelastic fluid droplet surrounded by a Newtonian fluid, (ii) a squirmer confined within a Newtonian fluid droplet surrounded by a viscoelastic fluid, and (iii) a squirmer confined within a viscoelastic fluid droplet surrounded by another viscoelastic fluid.

\subsection{Squirmer in a viscoelastic fluid droplet surrounded by a Newtonian fluid} \label{results1}

\vspace{0.5cm}

This case corresponds to De\(_2 = 0\); therefore, the regular perturbation approach is applied using De\(_1\). The \(\mathcal{O}(\text{De}_1)\) corrections to the pressure and flow fields, as well as the resulting swimming speeds, are derived following the method outlined in Sec.~\ref{sec:O_De_soln}.\\

Before exploring the effects of droplet confinement, we first consider the behaviour in the limit \( \alpha \to \infty \), which corresponds to a squirmer in an unbounded viscoelastic fluid domain characterized by De\(_1\).

\subsection*{Squirmer in an unbounded Oldroyd-B fluid ($\alpha \to \infty$)}

\vspace{0.5cm}

In the limit $\alpha \to \infty$, the $\mathcal{O}(\text{De}_1)$ correction to the swimming speed of the squirmer is given by,
\begin{equation}
  V_s^{\text{De}_1}|_{\alpha\to\infty} =    V_s^{\text{U, } \text{De}_1} = -\frac{2}{15}\beta m_1,
\end{equation}
which is consistent with previous studies \cite{datt2019note,binagia2020swimming}. Therefore, the total swimming speed is given by, $V_s|_{\alpha\to\infty} = V_s^\text{U} = \frac{2}{3}-\text{De}_1\frac{2}{15}\beta m_1$.\\

\begin{figure}[!tbh]
  \centering
 \includegraphics[width=0.4\linewidth]{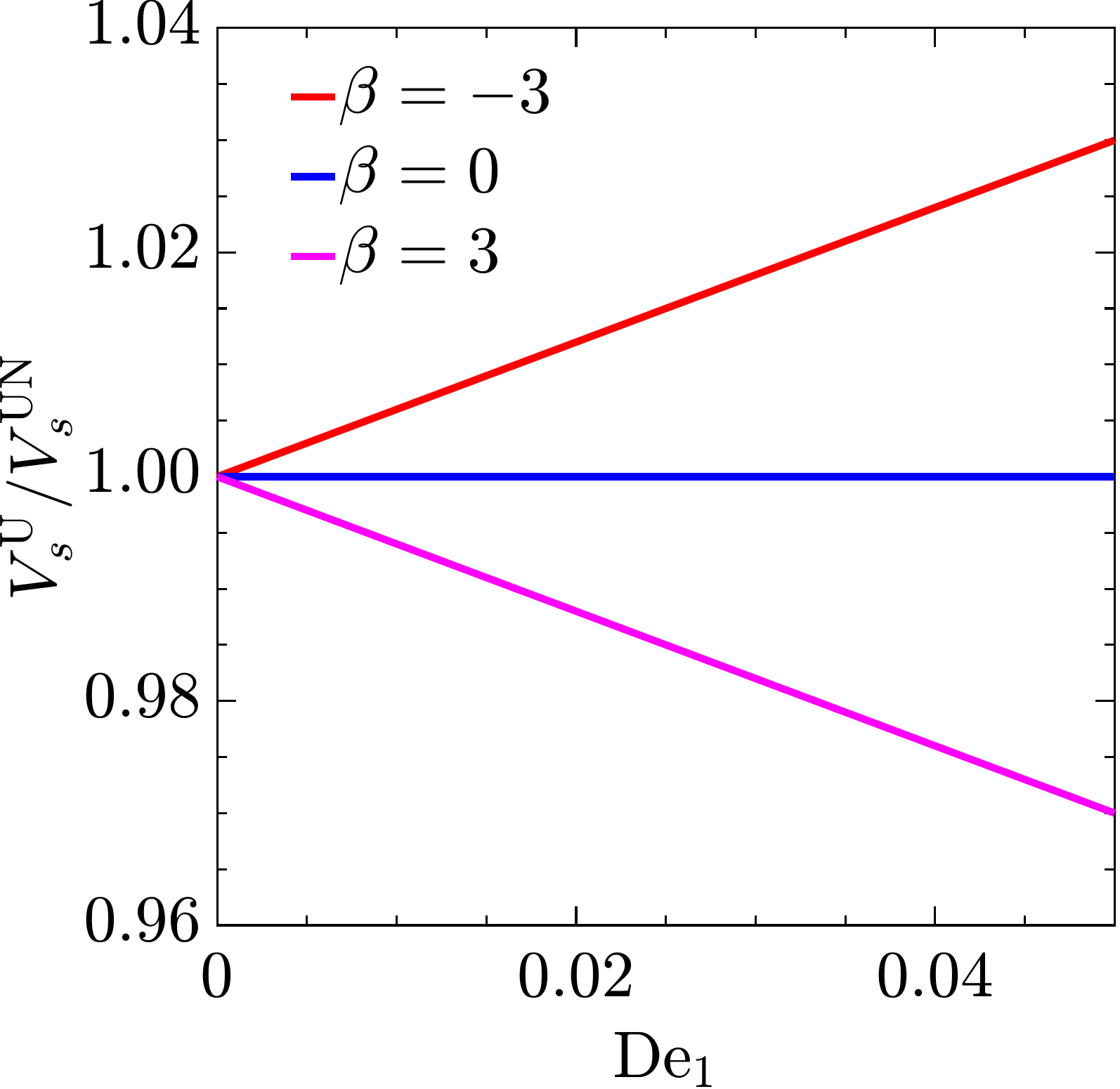}
  \caption{\label{fig:VsU_swimgspeed}  The normalised swimming speed \( V_s^\text{U}/V_s^\text{UN} \) of a squirmer in an unbounded Oldroyd-B fluid \kvs{($\alpha \to \infty$)} as a function of \( \text{De}_1 \) for \kvs{$m_1 = 1$} and various values of \( \beta \).}
\end{figure}

Figure~\ref{fig:VsU_swimgspeed} shows the variation in swimming speed \( V_s^\text{U} \) with respect to the Deborah number \( \text{De}_1 \) for different swimmer types, identified by the parameter \( \beta \). Notably, for a neutral swimmer (\( \beta = 0 \)), the swimming speed remains constant across varying values of \( \text{De}_1 \), showing the absence of \( \mathcal{O}(\text{De}_1) \) correction due to viscoelasticity. The swimming speed of a pusher (\( \beta = -3 \)) increases with increasing \( \text{De}_1 \), while for a puller (\( \beta = 3 \)), the swimming speed decreases as \( \text{De}_1 \) increases.\\

\begin{figure}[!tbh]
  \centering
 \includegraphics[width=\linewidth]{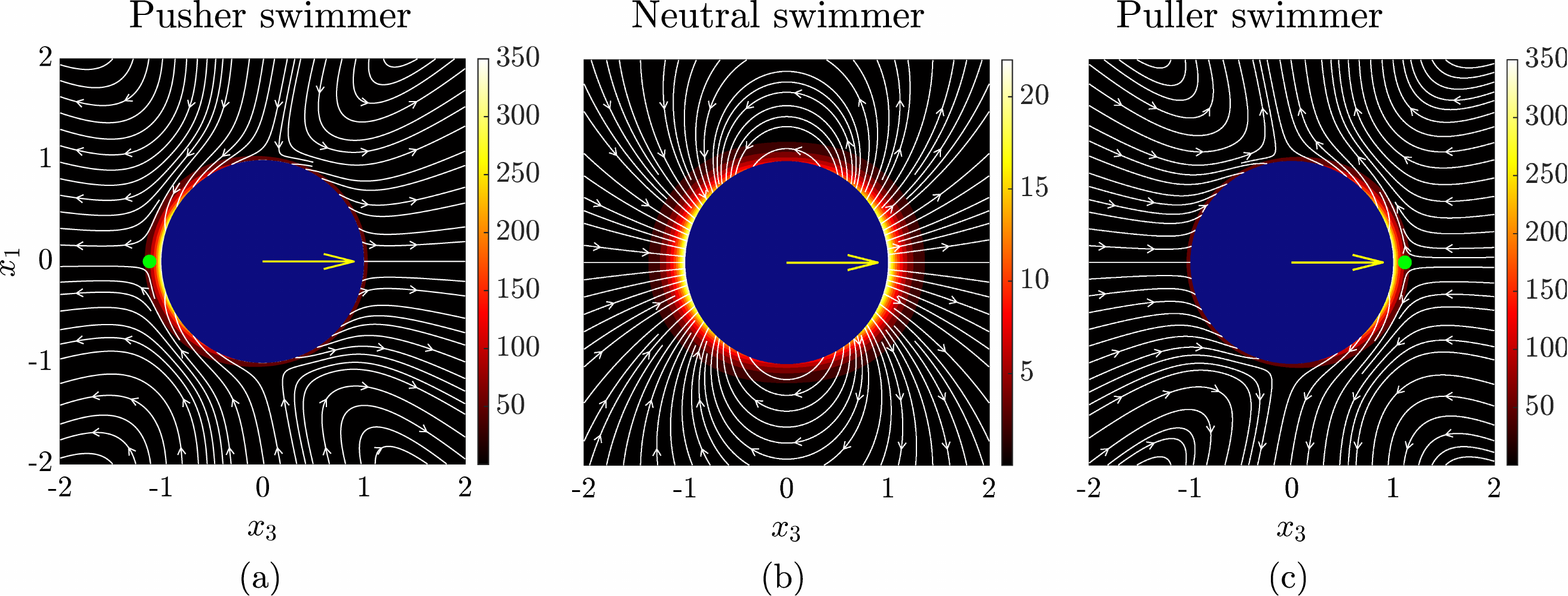}
  \caption{\label{fig:TrU} The trace of the polymeric stress tensor (shown by the colorbar, Eqn.~\ref{eqn:first_order_poly_st}) overlaid with the \(\mathcal{O}(1)\) velocity field, plotted as streamlines, for (a) a pusher swimmer ($\beta = -3$), (b) neutral swimmer ($\beta = 0$) and (c) puller swimmer ($\beta = 3$) in the limit $\alpha \to \infty$. The blue-filled circle indicates the squirmer, the yellow arrow indicates the orientation of the squirmer, and the green dot in (a) and (b) marks the stagnation point.}
\end{figure}

The observed differences in swimming speeds depicted in Fig.~\ref{fig:VsU_swimgspeed} can be understood by examining the polymeric stress induced by the \(\mathcal{O}(1)\) flow field (Eqn.~\ref{eqn:first_order_poly_st}). This stress, generated by the polymers in response to the \(\mathcal{O}(1)\) flow field, acts as a forcing term in the \(\mathcal{O}(\text{De}_1)\) governing equations (Eqn.~\ref{eqn:first_order_str}). A key indicator of the viscoelastic effects is the trace of the polymeric stress tensor, which is shown in Fig.~\ref{fig:TrU}, overlaid with the \(\mathcal{O}(1)\) velocity field for different values of \(\beta\). For a neutral swimmer (\(\beta = 0\)), the source-sink velocity field generates a fore-aft symmetric distribution of polymeric stress (Fig.~\ref{fig:TrU}(b)), which effectively cancels out, resulting in no change in swimming speed at  $\mathcal{O}( \text{De}_1)$. For a pusher swimmer (\(\beta = -3\)), the fluid is drawn toward the swimmer on both lateral sides and expelled along the anterior and posterior sides. The high polymeric stress is concentrated at the back of the swimmer~\cite{dwivedi2023deforming} near the stagnation point, where velocity gradients are strong (Fig.~\ref{fig:TrU} (a)). This stress effectively acts in the direction of the swimmer, providing an additional propulsive force that enhances the swimming speed. Conversely, for a puller swimmer (\(\beta = 3\)), the polymeric stress is concentrated at the front of the squirmer (Fig.~\ref{fig:TrU} (c)). This stress opposes the swimmer’s propulsion, effectively acting as a resistive force that reduces the swimming speed as \( \text{De}_1 \) increases.\\

Quantitatively, the forces acting on the squirmer at \(\mathcal{O}(\text{De}_1)\) (Eqn.~\ref{eqn:first_order_force_free}) can be calculated and conveniently decomposed into three contributions \cite{binagia2020swimming}: (i) force due to the polymeric stress defined by the \(\mathcal{O}(1)\) flow field (Eqn.~\ref{eqn:first_order_poly_st}) given by \( F_{poly}^{\text{De}_1} = \frac{8}{3}\beta m_1 \), (ii) force due to \(\mathcal{O}(\text{De}_1)\) pressure field, \( F_{pr}^{\text{De}_1} = -\frac{22}{5}\beta m_1 - V_s^{\text{U, }\text{De}_1} \), and (iii) force due to \(\mathcal{O}(\text{De}_1)\) viscous stresses \( F_{vi}^{\text{De}_1} = \frac{4}{3}\beta m_1 - 2 V_s^{\text{U, }\text{De}_1} \). To compare the relative importance of three contributions, we can consider the pumping problem for which \( V_s^{\text{U},\text{ De}_1}=0 \). Then, it is evident that the dominant viscoelastic contribution comes from the pressure force, \kvs{similar to the findings of \citet{neo2024effects}. This} pressure force is positive for pushers, enhancing their swimming speed, and negative for pullers, thereby reducing their swimming speed.

\subsubsection{Effect of droplet confinement}

\vspace{0.5cm}

\begin{figure}[!tbh]
  \centering
 \includegraphics[width=0.45\linewidth]{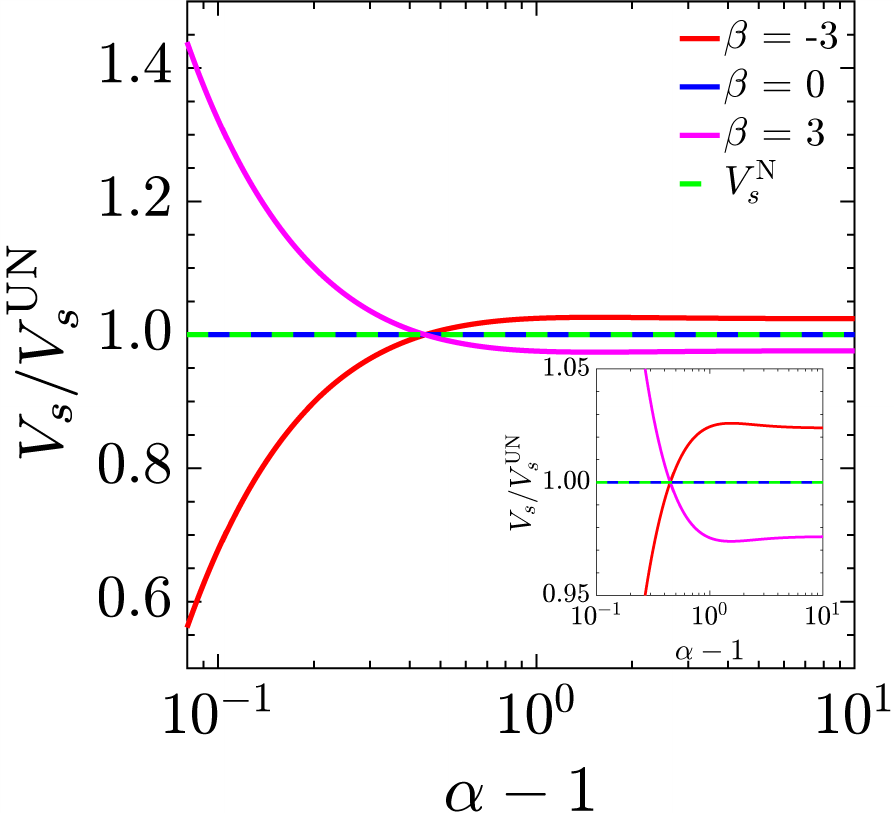}
  \caption{\label{fig:case1_swim_speeds} The normalized swimming speed of the squirmer, \(V_s/V_s^\text{UN}\), as a function of the thickness of the viscoelastic fluid layer, \(\alpha-1\), for different $\beta$. Here, the surrounding fluid is Newtonian and the droplet fluid is viscoelastic ($\text{De}_1 = 0.04$ and $m_1 = 1$),  $\lambda = 1$ .}
\end{figure}

Next, we relax the limit $\alpha \to \infty$, that of an unconfined squirmer, and examine the effect of droplet confinement on the swimming speed. This case corresponds to viscoelastic fluid present only in the vicinity of the squirmer, with a thickness of \(\alpha - 1\). Figure~\ref{fig:case1_swim_speeds} shows the impact of the thickness \(\alpha - 1\) of the viscoelastic layer on the swimming speed for various values of \(\beta\).\\

As discussed in section~\ref{sec:zeroeth_order}, when the droplet fluid is Newtonian and has the same viscosity as that of the surrounding fluid, \textit{i.e.,} \(\lambda = 1\), the swimming speed \(V_s^\text{N}\) of the squirmer is same as that of a squirmer in an unbounded Newtonian fluid, \(V_s^\text{UN}\), and remains independent of both \(\alpha\) and \(\beta\) \cite{reigh2017swimming}. Notably, even when the droplet fluid is viscoelastic, the swimming speed of the neutral swimmer (\(\beta = 0\)) remains unchanged, similar to the case of a Newtonian droplet fluid. However, when \(|\beta| > 0\), the swimming speed of the squirmer becomes dependent on \(\alpha\). As the confinement by the viscoelastic fluid increases, \textit{i.e.,} as the thickness of the viscoelastic fluid \((\alpha - 1)\) decreases, the swimming speed of pushers (\(\beta < 0\)) \kvs{initially} increases slightly compared to that of an unconfined swimmer \kvs{(Fig.~\ref{fig:case1_swim_speeds} inset)}, but then decreases significantly as the droplet fluid thickness becomes small. Conversely, for pullers, the swimming speed initially decreases slightly as the thickness of the droplet fluid decreases but increases as the confinement becomes more pronounced. \\


\begin{figure}[!tbh]
  \centering
  \includegraphics[width=\linewidth]{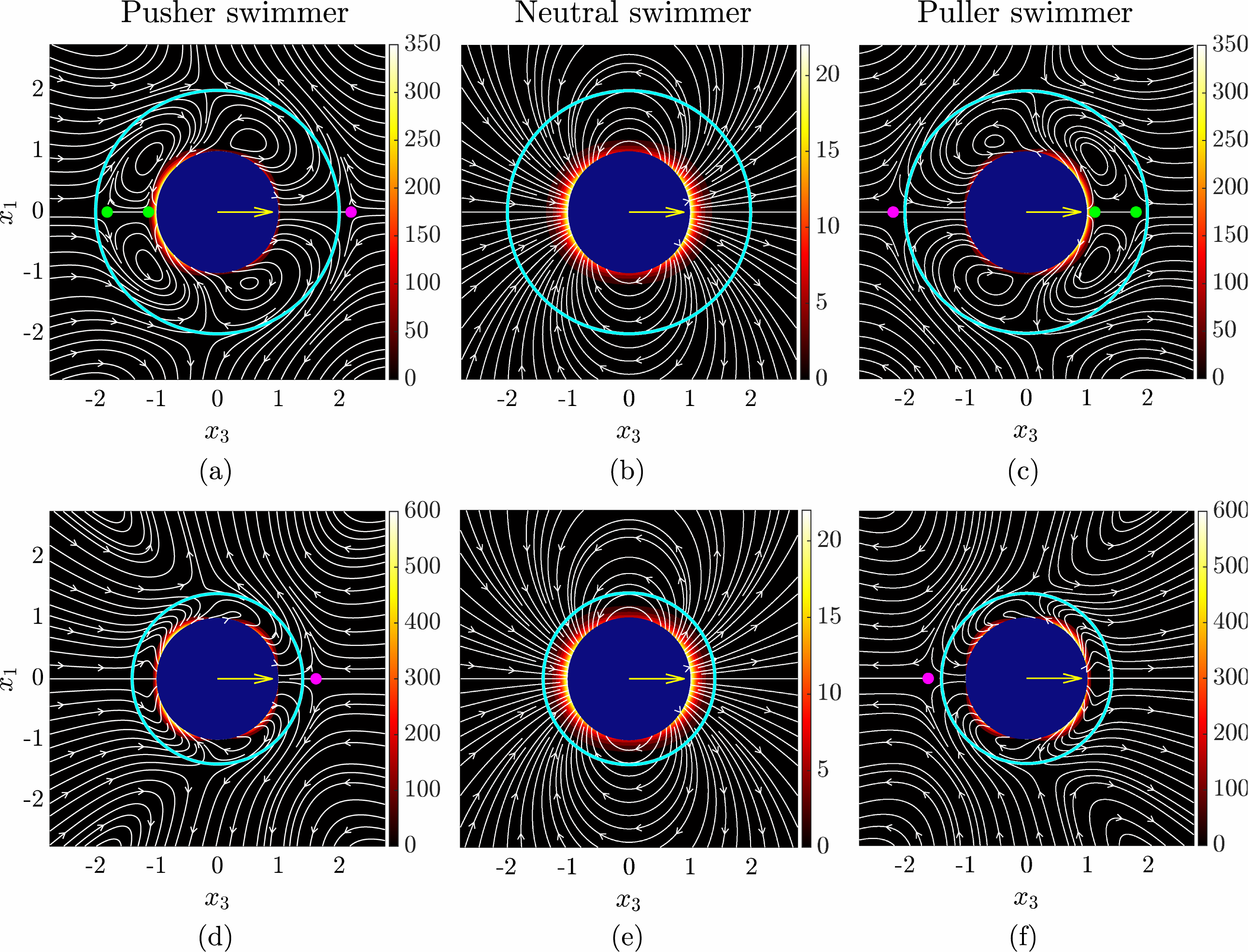}
  \caption{\label{fig:case1_FF} The trace of the polymeric stress tensor (shown by the colorbar) overlaid with the \(\mathcal{O}(1)\) velocity field of an active compound particle. The cyan circle represents the droplet interface, and the blue-filled circle indicates the squirmer, with its orientation marked by a yellow arrow. White lines with arrows are streamlines.  Stagnation points inside the droplet are marked in green, and those outside the droplet are marked in magenta. The top row (a-c) corresponds to a size ratio \(\alpha = 2\), and the bottom row (d-f) corresponds to a size ratio \(\alpha = 1.4\). The first column (a, d) represents pusher swimmers, the middle column (b, e) are neutral swimmers, and the last column (c, f) are puller swimmers.}
\end{figure}

This trend in change in the swimming speed with respect to variation in the size ratio can be understood by analyzing the polymeric stress induced by the \(\mathcal{O}(1)\) velocity field, similar to the discussion of the unbounded case presented in the previous section. Figure~\ref{fig:case1_FF} shows the velocity field and associated polymeric stress of an active compound particle for two different size ratios, \(\alpha = 2\) and \(\alpha = 1.4\). \\

When the size ratio \(\alpha = 2\), the flow field generated by the neutral swimmer and the corresponding polymeric stress remains unaltered (Fig.~\ref{fig:case1_FF}(b)) compared to that in an unbounded fluid. However, for both pusher (Fig.~\ref{fig:case1_FF}(a)) and puller (Fig.~\ref{fig:case1_FF}(c)) swimmers, the presence of the droplet interface leads to the formation of four distinct vortices, each located in one of the quadrants inside the droplet \cite{reigh2017swimming}. Additionally, along the axis of symmetry, the presence of droplet confinement introduces two additional stagnation points - one inside the droplet and one outside - adding to the one stagnation point near the squirmer surface that already existed in the unbounded case (Fig.~\ref{fig:TrU}). Specifically, for pushers, two stagnation points are located at the rear side of the squirmer inside the droplet, and one is at the front, outside the droplet (Fig.~\ref{fig:case1_FF}(a)). Conversely, for pullers, two stagnation points are located at the front side of the squirmer inside the droplet, and one is at the rear, outside the droplet (Fig.~\ref{fig:case1_FF}(c)). However, the flow field in the immediate vicinity of the squirmer closely resembles that of an unbounded squirmer, and the corresponding polymeric stress is also similar to that of an unbounded fluid for all values of \(\beta\) (Fig.~\ref{fig:case1_FF}(a)-(c)). As a result, the swimming speed of pushers and pullers, for large \(\alpha\) is not significantly altered compared to the swimming speed in an unbounded Oldroyd-B fluid (Fig.~\ref{fig:case1_swim_speeds}).\\

\begin{figure}[!tbh]
  \centering
  \includegraphics[width=0.7\linewidth]{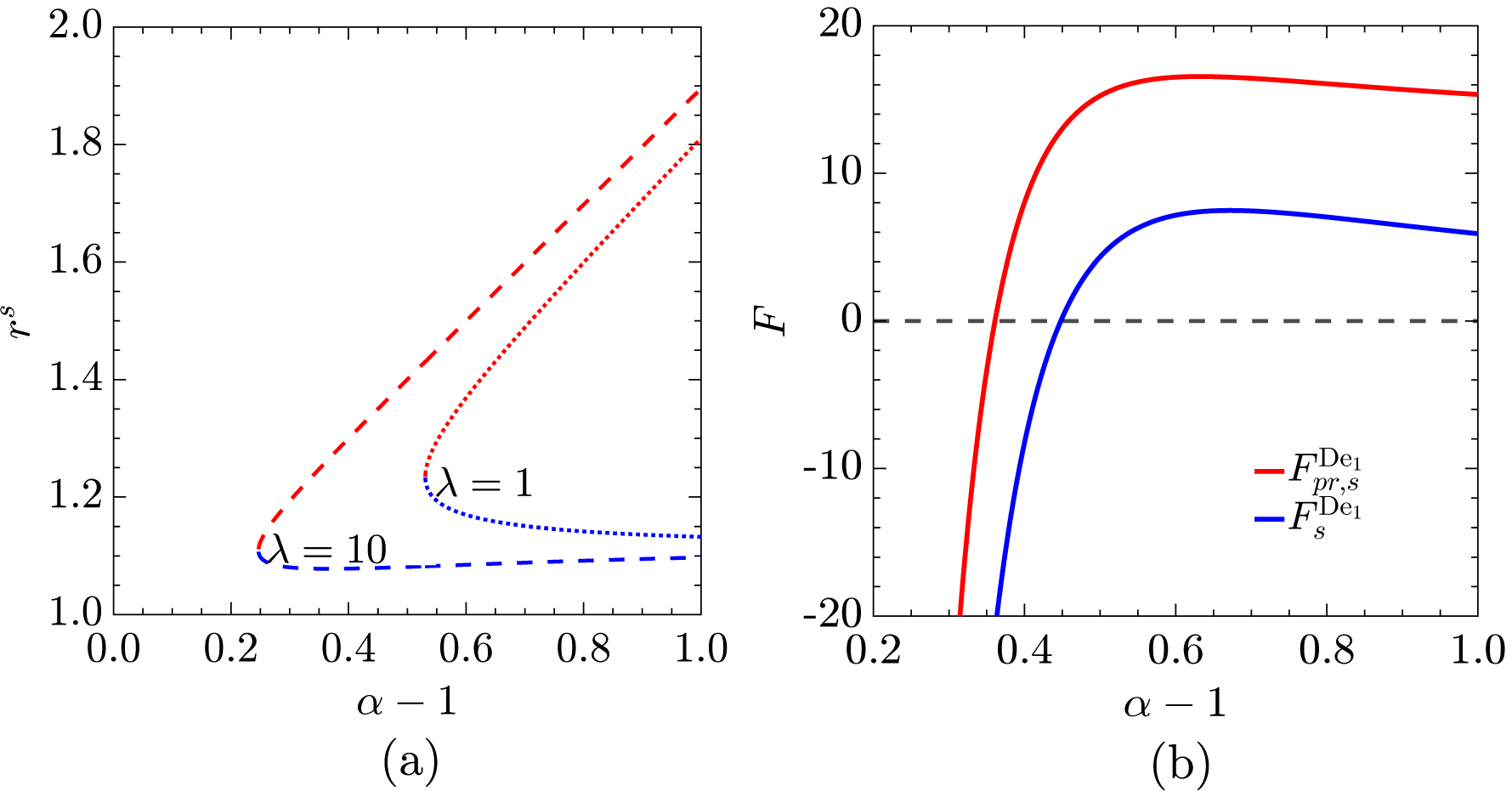}
  \caption{\label{fig:case1_stagp_forces} (a) Radial location (\(r^s\)) of the two stagnation points inside the fluid droplet (Fig.~\ref{fig:case1_FF} (a) \& (c)) as a function of the thickness of the viscoelastic fluid layer \(\alpha - 1\) for \(\beta = |3|\). The blue curves represent the stagnation point close to the squirmer surface, while the red curves represent the stagnation point away from the squirmer surface. (b) Force due to pressure (red curve) and total force (blue curve) on the squirmer in a fluid pumping problem as a function of  \(\alpha - 1\) for $\lambda = 1$, $\beta = -3$. The total force includes contributions from polymeric stress based on the \(\mathcal{O}(1)\) flow field, as well as \(\mathcal{O}(\text{De}_1)\) pressure and velocity fields. The dashed black line indicates the zero-force level.
}
\end{figure}

On the other hand, when the size ratio is smaller, such as \(\alpha = 1.4\), both the polymeric stress and the flow field exhibit significant changes for both pusher (Fig.~\ref{fig:case1_FF}(d)) and puller (Fig.~\ref{fig:case1_FF}(e)) swimmers. The flow field for a neutral swimmer remains unchanged (Fig.~\ref{fig:case1_FF}(f)). Specifically, in the case of pusher and puller swimmers, the polymeric stress increases as a result of stronger velocity gradients inside the confinement. Additionally, the vortices inside the droplet vanish, and no stagnation points are observed along the axis of symmetry within the droplet. This is clearly evident in Fig.~\ref{fig:case1_stagp_forces}, which shows the variation in the radial location of the two stagnation points inside the droplet with a change in the thickness of the viscoelastic fluid layer, \(\alpha-1\). As the thickness decreases, the two stagnation points move toward each other, eventually merging into a single point and then vanishing entirely. This transition in the flow field with the size ratio occurs because, as the size ratio decreases, the difference in swimming speeds between the droplet and the squirmer reduces, leading to more uniform fluid motion \cite{reigh2017swimming}. Note that the stagnation point and flow field outside the droplet remain unchanged.\\

\kvs{These transitions in the flow field inside the droplet are crucial, as they directly impact the polymeric stress distribution, which in turn influences the forces acting on the squirmer. A major contribution to the \(\mathcal{O}(\text{De}_1)\) force comes from the pressure field. Figure~\ref{fig:case1_PF} shows the \(\mathcal{O}(\text{De}_1)\) pressure field for \(\alpha = 2.0\) and \(1.4\). When the size ratio is large (Fig.~\ref{fig:case1_PF}(a)), the pressure is higher at the rear of the pusher than at the front, resulting in an increase in swimming speed. However, as the size ratio decreases (Fig.~\ref{fig:case1_PF}(d)), the pressure at the front increases due to external stagnation, countering the propulsive force induced by the rear stagnation points.} This is evident in Fig.~\ref{fig:case1_stagp_forces}(b), which shows the pressure and total forces acting on the squirmer in the fluid pumping problem (\(V_s^{\text{De}_1} = 0\)) as a function of the size ratio, for \(\lambda = 1\) and \(\beta = -3\). As the size ratio decreases, the pressure force initially increases, then decreases, and eventually becomes negative. The total force on the squirmer also exhibits a similar variation with the size ratio (Fig.~\ref{fig:case1_stagp_forces}(b) - blue curve). This trend mirrors the variation in the swimming speed with the size ratio for pushers depicted in Fig.~\ref{fig:case1_swim_speeds}.\\

The non-monotonic variation of the force on the squirmer with the thickness of the viscoelastic fluid layer, which determines the swimming speed depicted in Fig.~\ref{fig:case1_swim_speeds}, can be understood as follows: As the size ratio decreases, the polymeric stress inside the droplet increases due to stronger velocity gradients, leading to an initial increase in the force. However, as the size ratio continues to decrease, the two stagnation points on the rear side of the pusher swimmer vanish due to the decrease in the relative speed between the squirmer and the droplet (Fig.~\ref{fig:case1_FF} (d)). Simultaneously, the influence of the stagnation point outside the droplet increases, resulting in a reversal of the force and causing the observed decline in the velocity. This explains the dynamics for pushers. A complementary phenomenon occurs for pullers, where the trends in forces and swimming speeds are also mirrored.\\

\begin{figure}[!tbh]
  \centering
 \includegraphics[width=\linewidth]{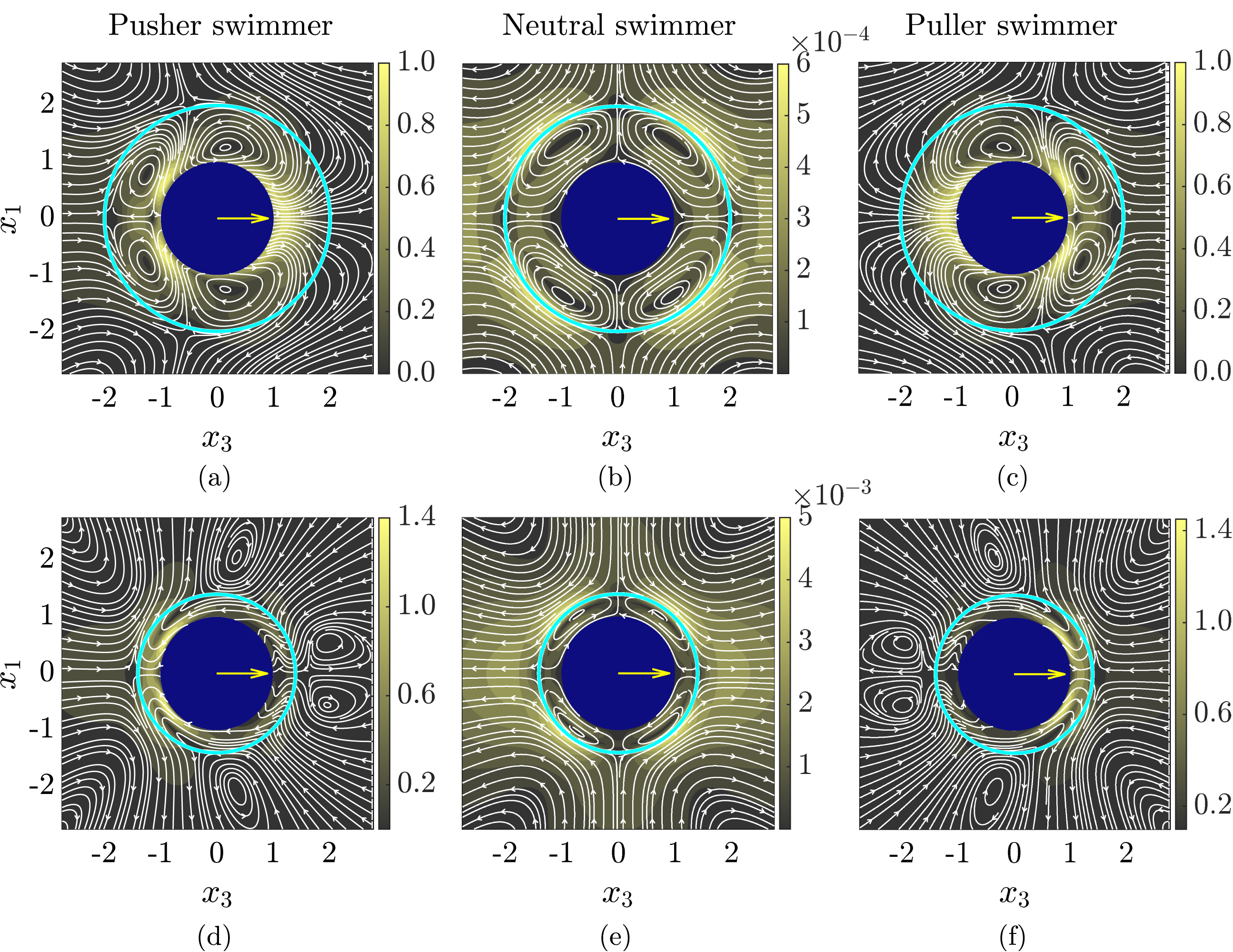}
  \caption{\label{fig:ODe_case1_FF} \(\mathcal{O}(\text{De}_1)\) correction to the velocity field of an active compound particle when the droplet fluid is viscoelastic and the surrounding fluid is Newtonian. White lines with arrows indicate streamlines, and the background colour corresponds to the magnitude of the velocity field. The top row (a-c) corresponds to a size ratio \(\alpha = 2\), and the bottom row (d-f) corresponds to a size ratio \(\alpha = 1.4\). The columns represent (a, d) pusher swimmers, (b, e) neutral swimmers, and (c, f) puller swimmers. The cyan circle represents the droplet interface, and the blue-filled circle indicates the squirmer, with its orientation marked by a yellow arrow. Here, $\lambda = 1.0$ and $m_1 = 1.0$.}
\end{figure}

Next, we examine the \(\mathcal{O}(\text{De}_1)\) correction to the velocity field to understand the influence of droplet viscoelasticity on the fluid flow. In Fig.~\ref{fig:ODe_case1_FF}, the velocity field is depicted for different swimmer types at two different size ratios, \(\alpha = 2\) and \(\alpha = 1.4\). When \(\alpha = 2\), the \(\mathcal{O}(\text{De}_1)\) correction for the neutral swimmer (\(\beta = 0\)) exhibits fore-aft and up-down symmetry (Fig.~\ref{fig:ODe_case1_FF}(b)), indicating that viscoelasticity does not introduce any net force at this order, resulting in no change to the swimming speed. For pushers (Fig.~\ref{fig:ODe_case1_FF}(a)), the \(\mathcal{O}(\text{De}_1)\) velocity field predominantly aligns with the direction of the \(\mathcal{O}(1)\) velocity, which correlates with the observed increase in swimming speed at larger \(\alpha\). Notably, there is also a maximum in the fluid velocity at the immediate front of the swimmer and a minimum at the immediate back on the axis of symmetry. In contrast, for pullers (Fig.~\ref{fig:ODe_case1_FF}(c)), the \(\mathcal{O}(\text{De}_1)\) correction opposes the direction of the \(\mathcal{O}(1)\) velocity field, consistent with the observed decrease in the swimming speed for large \(\alpha\). The maximum in the flow field is located at the back of the swimmer, while the minimum is at the front, on the axis of symmetry.\\

When the size ratio is small (\(\alpha = 1.4\)), the \(\mathcal{O}(\text{De}_1)\) correction in the velocity field of the neutral swimmer (Fig.~\ref{fig:ODe_case1_FF}(e)) retains a symmetric structure. However, for non-zero \(\beta\), the effects of viscoelasticity become more pronounced compared to the case with \(\alpha = 2\), due to the increased polymeric stress. For pushers (Fig.~\ref{fig:ODe_case1_FF}(d)), the maxima in the fluid velocity at the front disappear and instead appear on the rear side. This shift corresponds to the reversal of the force acting on the squirmer, as well as the observed decrease in swimming speed with decreasing \(\alpha\). Similarly, for pullers (Fig.~\ref{fig:ODe_case1_FF}(f)), the maxima in the flow at the rear side disappear and instead appear on the front side, consistent with the observed increase in swimming speed as \(\alpha\) decreases.  \\


\begin{figure}[!tbh]
  \centering
 \includegraphics[width=0.4\linewidth]{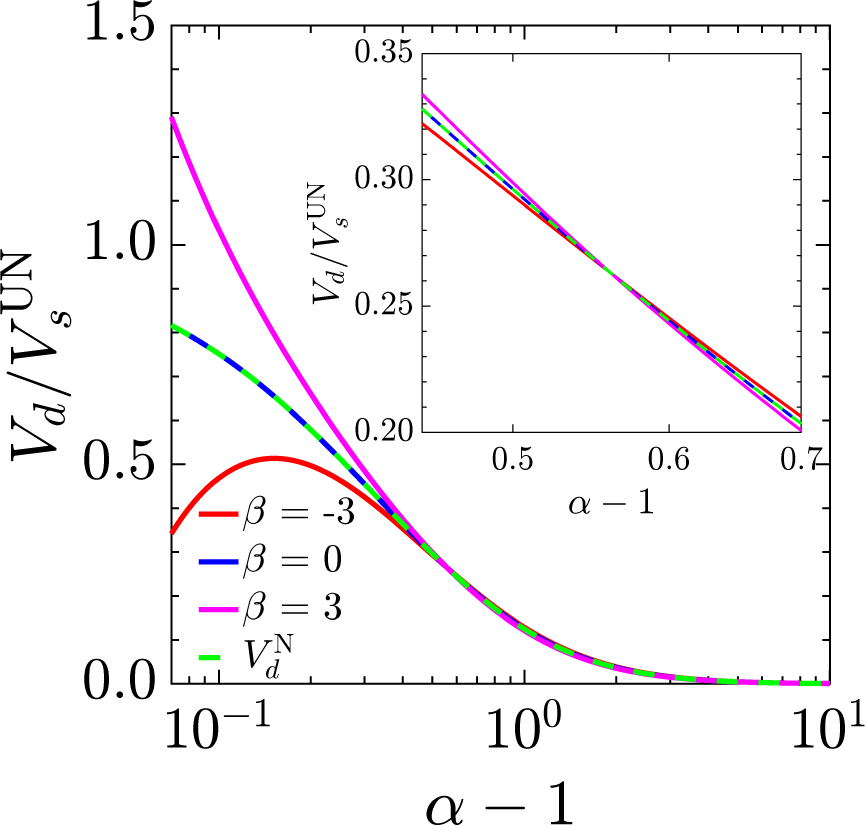}
  \caption{\label{fig:dropspeedcase1}  The normalised droplet speed \( V_d/V_s^\text{UN} \), as a function of the thickness of the viscoelastic fluid layer, \(\alpha-1\), for $\lambda = 1$, with varying $\beta$. Here, the droplet fluid is viscoelastic ($\text{De}_1 = 0.04$, $m_1 = 1.0$), and the surrounding fluid is Newtonian.}
\end{figure}

The self-propulsion of the encapsulated swimmer and the associated flow generated impart velocity to the confining droplet. Figure~\ref{fig:dropspeedcase1} depicts the dependence of the droplet speed on the thickness of the viscoelastic fluid layer, \(\alpha - 1\), for various values of \(\beta\). Notably, irrespective of \(\beta\), for large \(\alpha\), the droplet speed approaches zero, indicating the weakened hydrodynamic interactions as the droplet becomes significantly larger than the swimmer. For a neutral swimmer (\(\beta = 0\)), the droplet speed matches that of the squirmer confined within a Newtonian fluid droplet surrounded by a Newtonian fluid, similar to the observation that swimming speed \(V_s\) remains unchanged regardless of \(\alpha\) for \(\beta = 0\) (Fig.~\ref{fig:case1_swim_speeds}).
For pushers, the droplet speed exceeds that of the Newtonian counterpart (\(V_d^\text{N}\)) for intermediate \(\alpha-1\) (Fig.~\ref{fig:dropspeedcase1} inset), which parallels the initial increase in swimmer speed as the size ratio decreases. However, with further decrease in $\alpha-1$, the droplet speed decreases and gets lower than $V_d^\text{N}$, mirroring the decrease in swimmer speed observed for small droplet fluid thickness (Fig.~\ref{fig:case1_swim_speeds}). For pullers (\(\beta > 0\)), the droplet speed is lower than that of the Newtonian case at intermediate \(\alpha-1\). With the further decrease in the size ratio, the droplet speed increases and eventually exceeds \(V_d^\text{N}\) (Fig.~\ref{fig:dropspeedcase1} inset), mirroring the variation in the speed of a puller with the size ratio.
\\

\begin{figure}[!tbh]
  \centering
 \includegraphics[width=0.7\linewidth]{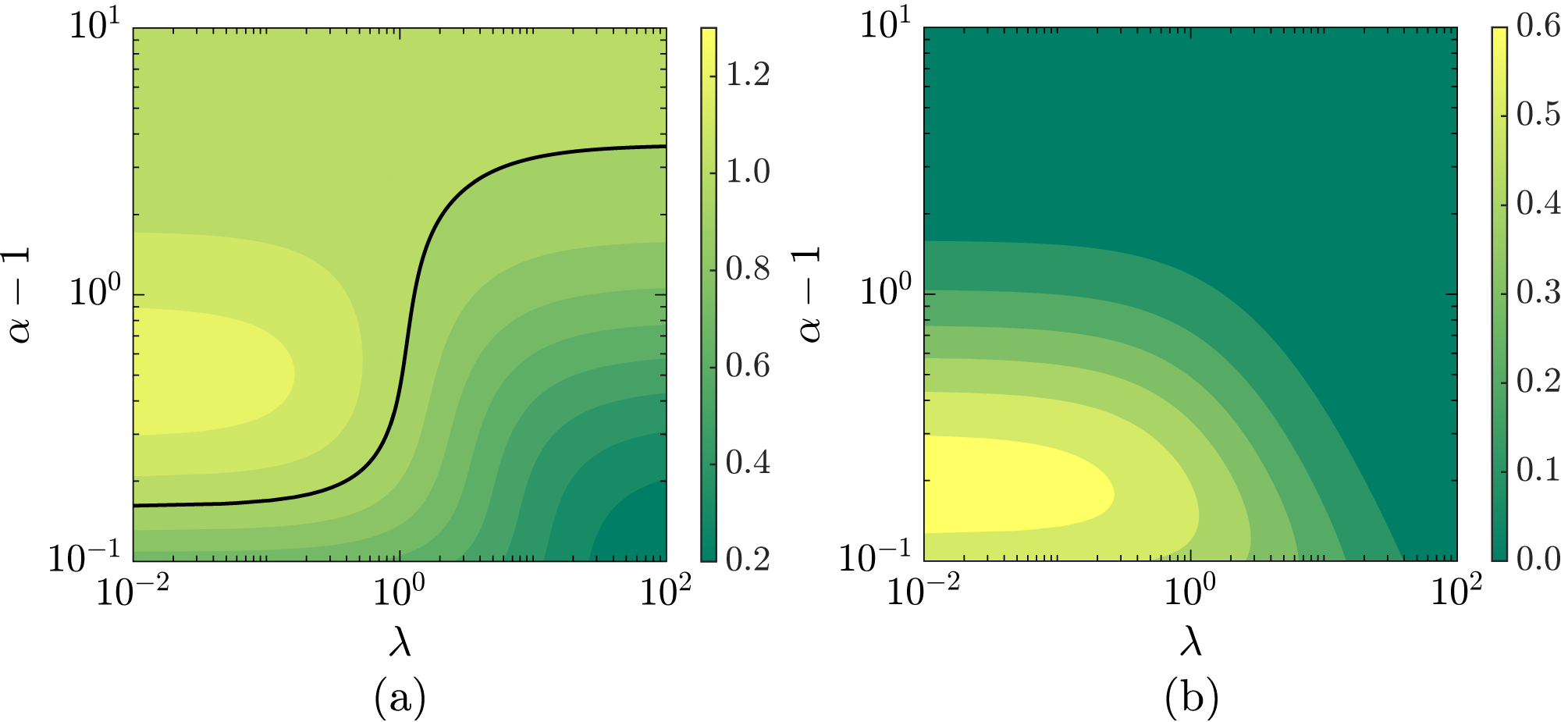}
  \caption{\label{fig:case1phasediag}  The dependency of the normalized speed of (a) the pusher swimmer,  \(V_s/V_s^\text{UN}\), and (b) the droplet, \(V_d/V_s^\text{UN}\), on the thickness of the viscoelastic fluid layer \(\alpha - 1\) and the viscosity ratio \(\lambda\) for $\beta = -3$. Here, the droplet fluid is viscoelastic ($\text{De}_1 = 0.04$, $m_1$= 1.0), and the surrounding fluid is Newtonian. The solid black line indicates where the normalized swimming speed equals one, signifying no change in speed compared to the unbounded Newtonian fluid case. The dashed black line marks the region where the swimming speed is zero, indicating a reversal in the swimming speed.}
\end{figure}

Lastly, we examine the impact of the viscosity ratio on the speed of the swimmer and the confining droplet. Given the complementary behaviours of pushers and pullers, we focus the discussion hereafter on the pusher swimmer. Figure~\ref{fig:case1phasediag} illustrates the dependency of the normalized speed of a swimmer, \(V_s/V_s^{\text{UN}}\) (Fig.~\ref{fig:case1phasediag}(a)), and the droplet, \(V_d/V_s^{\text{UN}}\) (Fig.~\ref{fig:case1phasediag}(b)), on the thickness of the viscoelastic fluid layer, \(\alpha - 1\), and the viscosity ratio, \(\lambda\), for \(\text{De}_1 = 0.04\) and \(\beta = -3\). For a given value of $\alpha-1$, the swimmer's speed decreases with the viscosity ratio. However, for large values of the size ratio, the swimmer's speed is independent of the viscosity ratio. On the other hand, for small \(\lambda\), the swimmer speed shows a non-monotonic variation with \(\alpha-1\) with a maximum at intermediate \(\alpha-1\), in agreement with the Fig.~\ref{fig:case1_swim_speeds}. In contrast, for large \(\lambda\), the swimmer speed increases monotonically with \(\alpha\). This occurs because, as \(\lambda\) increases, the flow outside the droplet becomes weaker, and the droplet interface approaches a no-slip boundary condition. Consequently, the stagnation points inside the droplet are maintained over a large range of size ratios (Fig.~\ref{fig:case1_stagp_forces}(a)). The speed of the droplet, \(V_d\), follows a similar trend to that of the swimmer speed (Fig.~\ref{fig:case1phasediag}(b)). Moreover, the droplet speed is generally lower than the swimmer's speed, indicating that the concentric configuration is not stable but instantaneous. This nuanced interplay between the viscosity ratio and size ratio underscores the complex dynamics governing the swimmer's propulsion within a viscoelastic droplet surrounded by a Newtonian fluid. \\

\subsection{Squirmer in a Newtonian fluid droplet surrounded by a viscoelastic fluid} \label{results2}

Next, we discuss Case 2, where the squirmer is confined in a Newtonian fluid droplet surrounded by a viscoelastic fluid. This case corresponds to \(\gamma = 0\), as outlined in Sec.~\ref{sec:O_De_soln}. 

\begin{figure}[!tbh]
  \centering
 \includegraphics[width=\linewidth]{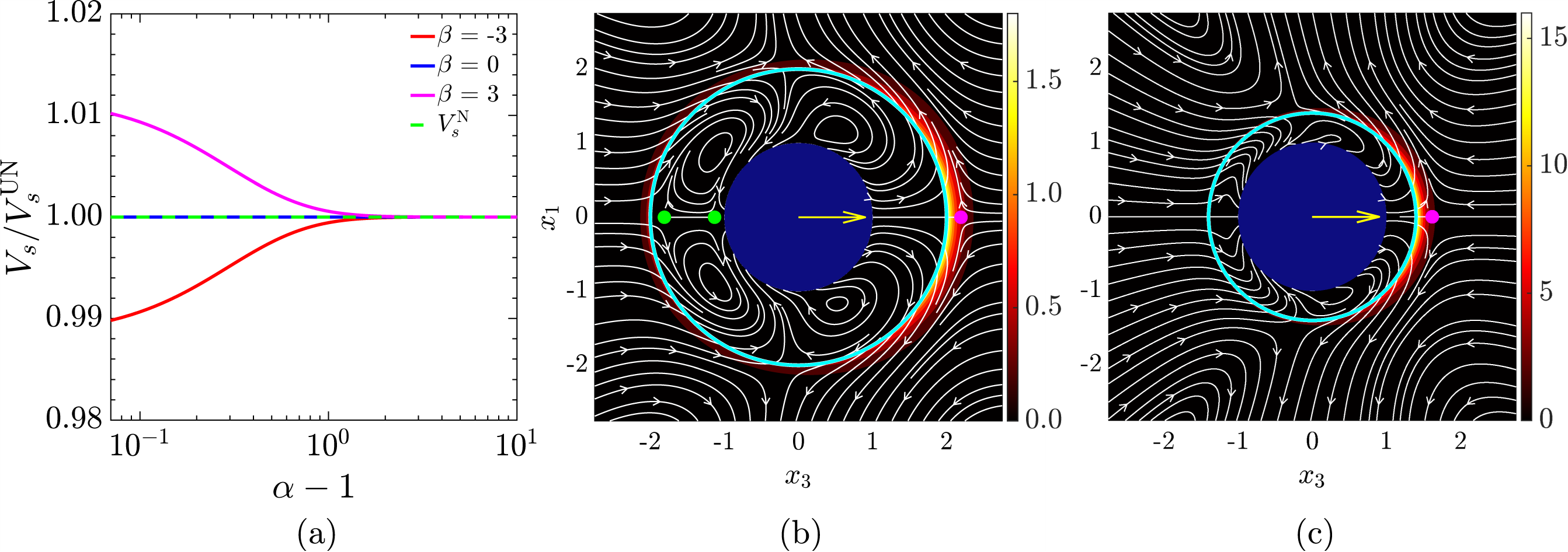}
  \caption{\label{fig:case2_plots} (a) The normalized swimming speed of the squirmer, \(V_s/V_s^\text{UN}\), plotted against the thickness of the droplet fluid, \(\alpha - 1\), with varying \(\beta\), for \(\lambda = 1\),  \(\gamma = 0\), and \(\text{De}_2 = 0.04\) -  for a system where the droplet fluid is Newtonian, and the surrounding fluid is viscoelastic. (b) and (c) show the trace of the polymeric stress tensor (represented by the colorbar) overlaid with the \(\mathcal{O}(1)\) velocity field. The parameters are \(\alpha = 2.0\) and \(\alpha = 1.4\), respectively, with \(\beta = -3\) and \(\lambda = 1\). The cyan circle represents the droplet interface, and the blue-filled circle indicates the squirmer, with its orientation marked by a yellow arrow. Stagnation points inside the droplet are marked in green, and those outside the droplet are marked in magenta.}
\end{figure}

Figure~\ref{fig:case2_plots}(a) illustrates the variation in the swimming speed of a squirmer as a function of the droplet thickness \(\alpha - 1\) for different values of \(\beta\), with the viscosity ratio \(\lambda = 1\), and \(\text{De}_2 = 0.04\). Irrespective of \(\beta\), the swimming speed of the squirmer matches that of it in an unbounded Newtonian fluid (\(V_s^{\text{UN}}\)) when \(\alpha\) is large. However, the \(\mathcal{O}(\text{De}_2)\) correction to the swimming speed in the limit \(\alpha \to 1\) is given by $V_s^{\text{De}_2}|_{\alpha\to1} = \frac{\beta m_2}{15}$. Interestingly, the swimming speed in this limit is not the same as that of a squirmer in an unbounded Oldroyd-B fluid. Therefore, as the droplet thickness decreases, the swimming speed of pusher swimmers decreases and eventually saturates at a value lower than that in an unbounded Newtonian fluid.\\

This reduction in swimming speed can be understood by examining the polymeric stress distributions shown in Fig.~\ref{fig:case2_plots}(b) and (c) for \(\alpha = 2\) and \(\alpha = 1.4\), respectively. Unlike in Case 1, where the stagnation points inside the droplet are of primary relevance, here the stagnation point outside the droplet plays a crucial role. For pushers, this stagnation point is located on the front side of the droplet where the polymeric stress is high. Consequently, the increased polymeric stress in the direction of motion results in a reduction of the swimming speed, similar to the behaviour observed for pullers in Case 1. However, when \(\alpha\) is large, the effect of the polymeric stress in the surrounding viscoelastic fluid on the squirmer's dynamics is negligible, allowing the swimmer to maintain a speed similar to that in an unbounded Newtonian fluid. Henceforth, as \(\alpha\) decreases and the droplet becomes thinner, the interaction between the squirmer and the surrounding viscoelastic fluid enhances, leading to an increase in polymeric stress and a subsequent reduction in swimming speed. \kvs{Additionally, the force on the squirmer due to the polymeric stress generated by the \(\mathcal{O}(1)\) flow field (\( F_{poly}^{\text{De}_2} \)) is zero in this case. This is because the squirmer is surrounded by a Newtonian droplet fluid and does not come into direct contact with the viscoelastic fluid surrounding the droplet. Therefore, the polymeric stress in the surrounding fluid influences the squirmer dynamics solely through the forces arising from \(\mathcal{O}(\text{De}_2)\) pressure and viscous stresses (see~\ref{sec:AppendixE}).}

\subsection{Squirmer in a viscoelastic fluid droplet surrounded by a viscoelastic fluid} \label{results3}

\begin{figure}[!tbh]
  \centering
 \includegraphics[width=\linewidth]{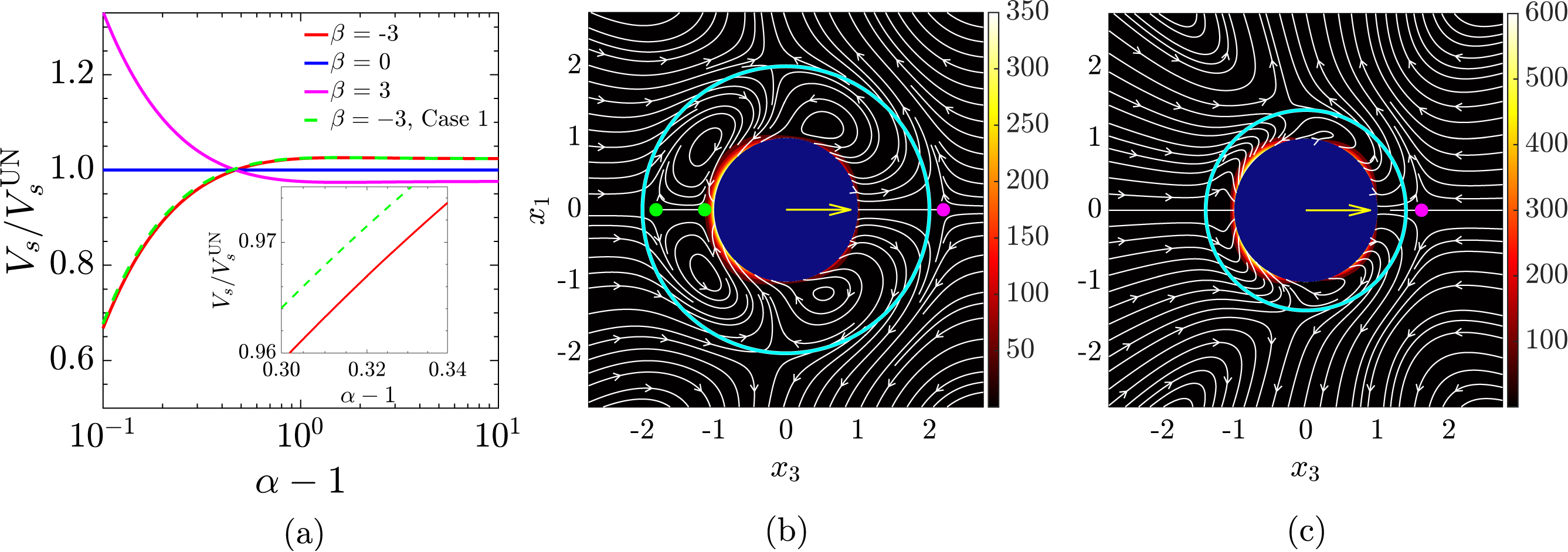}
  \caption{\label{fig:case3_plots}  (a) The normalized swimming speed of the squirmer, \(V_s/V_s^\text{UN}\), plotted against the thickness of the droplet fluid, \(\alpha - 1\), for \(\lambda = 1\), with varying \(\beta\), \(\gamma = 1\), and \(\text{De}_2 = 0.04\). (b) and (c) show the trace of the polymeric stress tensor (represented by the colorbar) overlaid with the \(\mathcal{O}(1)\) velocity field for a system where both the droplet and surrounding fluids are viscoelastic. The parameters are \(\alpha = 2.0\) and \(\alpha = 1.4\), respectively, with \(\beta = -3\) and \(\lambda = 1\). The cyan circle represents the droplet interface, and the blue-filled circle indicates the squirmer, with its orientation marked by a yellow arrow. Stagnation points inside the droplet are marked in green, and those outside the droplet are marked in magenta.}
\end{figure}

Lastly, we discuss Case 3, where the squirmer is confined in a viscoelastic fluid droplet surrounded by a viscoelastic fluid (\(\gamma = 1\)). Figure~\ref{fig:case3_plots}(a) illustrates the variation in the swimming speed of a squirmer as a function of the droplet thickness \(\alpha - 1\) for different values of \(\beta\). The variation in the swimming speed of a pusher swimmer with the thickness of the droplet fluid closely mirrors the behaviour observed in Case 1, where the droplet fluid is viscoelastic, and the surrounding fluid is Newtonian. However, the swimming speed in the present case is slightly lower compared to that in Case 1 (Fig,~\ref{fig:case3_plots} inset). This is because, the viscoelasticity of the surrounding fluid generates polymeric stress at the stagnation point on the front side of the droplet, which acts to decrease the swimming speed. However, the polymeric stress outside the droplet is significantly lower than the polymeric stress inside the droplet (Fig.~\ref{fig:case3_plots}(b) \& (c)), thereby having a minimal effect on the overall swimming speed.\\

The similarity between Cases 1 and 3 suggests that the properties of the surrounding fluid, whether Newtonian or viscoelastic, play a relatively minor role in determining the swimming speed when $m_1 = m_2$. This is because, in the limit $\alpha \to \infty$, the dynamics are dominated by the inner droplet fluid (Fig.~\ref{fig:case3_plots}(b)). \kvs{Specifically, in this limit, the configuration reduces to that of a squirmer in an unbounded viscoelastic droplet fluid, and the role of the outer fluid becomes minimal.} Conversely, when $\alpha$ is small, the polymeric stress within the droplet fluid becomes extremely large and outweighs the polymeric stress in the outer fluid as shown in Fig.~\ref{fig:case3_plots}(c), and predominantly governs the motion of the squirmer. Thus, in both limits, the droplet fluid's viscoelasticity is the dominant factor.

\section{Conclusions} \label{conclu}

In this study, we investigated the hydrodynamics of an active compound particle, a squirmer confined within a viscoelastic fluid droplet that is suspended in another viscoelastic medium. By assuming small Deborah and Capillary numbers, we derived the \(\mathcal{O}(\text{De})\) correction to the velocity field and examined the corresponding swimming speeds of both the squirmer and the fluid droplet. We examined the effects of fluid viscoelasticity in three distinct cases: (i) a viscoelastic fluid droplet in a Newtonian fluid, (ii) a Newtonian fluid droplet in a viscoelastic fluid, and (iii) a viscoelastic fluid droplet in another viscoelastic fluid.\\

In Case 1, we examined a squirmer within a viscoelastic droplet suspended in a Newtonian fluid. In the limit of large size ratio of the droplet to the squirmer, the results are consistent with findings in the literature \cite{de2015locomotion, datt2019note}: pushers tend to swim faster, pullers swim slower, and there is no \(\mathcal{O}(\text{De})\) correction to the swimming speed of a neutral swimmer. This behaviour is explained by analyzing the polymeric stress generated by the \(\mathcal{O}(1)\) velocity field. For neutral swimmers, the stress distribution is fore-aft symmetric, leading to no change in swimming speed. In the case of pushers, high polymeric stress is located on the rear side of the squirmer near a stagnation point, enhancing their swimming speed. Conversely, for pullers, the stagnation point and high polymeric stress are located at the front, opposing their motion and reducing their swimming speed.\\

We further explored how the droplet confinement alters the influence of viscoelasticity on squirmer dynamics. For pushers, confinement initially enhances the swimming speed as the polymeric stresses near the rear stagnation point within the droplet induce additional propulsive forces. However, as the confinement strengthens (droplet size decreases), the stagnation points within the droplet vanish, and the external stagnation point, located at the front of the squirmer, plays a more significant role. This change causes a reduction in the propulsive force on the squirmer, leading to a decrease in the swimming speed. Conversely, for pullers, confinement initially decreases the swimming speed due to resistive forces. As the internal stagnation points vanish, the external stagnation point at the rear acts to enhance propulsion, leading to an increase in the swimming speed as the droplet size decreases. Additionally, an increase in the viscosity ratio of the surrounding fluid to the droplet fluid acts to decrease the swimming speed except for large size ratios. Furthermore, the speed of the droplet follows a similar trend to that of the swimmer's speed.\\

In Case 2, where a squirmer is confined within a Newtonian droplet suspended in a viscoelastic fluid, the external stagnation point plays a crucial role in determining the microswimmer dynamics. For pushers, the stagnation point is located on the front side of the droplet, causing the swimming speed to decrease as confinement strengthens, eventually saturating at a constant value lower than that in a purely Newtonian fluid. Conversely, for pullers, the external viscoelastic fluid generates additional propulsive forces through the stagnation point at the rear of the droplet. As a result, the swimming speed increases with decreasing confinement. These findings suggest that a thin coating of Newtonian fluid on a squirmer in a viscoelastic fluid can enhance the speed of pullers while decreasing the speed of pushers. On the other hand, in Case 3, where the squirmer is confined within a viscoelastic fluid droplet surrounded by another viscoelastic fluid, the swimming dynamics closely resemble those observed in Case 1, where the droplet is viscoelastic and the surrounding fluid is Newtonian. This similarity suggests that when both the droplet and the surrounding fluids are viscoelastic, the surrounding fluid's viscoelastic properties play a relatively minor role compared to the dominant influence of the droplet fluid's viscoelasticity.\\

\kvs{Lastly, we note that our analysis is based on a concentric configuration, and the insights gained can also be useful in weakly eccentric configurations \cite{shaik2018locomotion}. Future research could explore the effects of viscoelasticity on eccentric configurations of microswimmers. Specifically, it would be interesting to investigate whether there is a non-zero viscoelastic correction to the swimming speed for the neutral swimmer in eccentric configurations due to the broken fore-aft symmetry in the configuration.} Future research may also extend the present investigations by exploring the \(\mathcal{O}(\text{Ca})\) corrections to account for droplet deformation~\cite{chaithanya2020deformation,chaithanya2023active}. Additionally, investigating the combined effects of viscoelasticity and droplet deformation through \(\mathcal{O}(\text{De} \cdot \text{Ca})\) corrections could reveal more complex interactions between these two factors. Experiments with synthetic swimmers confined in fluid droplets could effectively test and validate the theoretical predictions, offering valuable insights for practical applications. In cases with a larger Deborah number, where stronger viscoelastic effects occur, \kvs{semi-analytical approaches such as weak-coupling expansion~\cite{moore2012weak} or} numerical simulations will be essential to investigate dynamics beyond the current analytical framework.

\appendix
\section{Equations determining the constants in Eqn.~\ref{eqn:zeroth_str_sol_1} \&~\ref{eqn:zeroth_str_sol_2}}
\label{sec:AppendixA}

\begin{align}
&V_s^\text{N} + A_{11}^\text{N}  + B_{11}^\text{N}  + C_{11}^\text{N}  + D_{11}^\text{N}  = 0 \label{eq:1}\\
&2 V_s^\text{N} + 4 A_{11}^\text{N}  + 2 B_{11} ^\text{N} + C_{11} ^\text{N} = 2 + D_{11} ^\text{N} \label{eq:2}\\
&V_d^\text{N} + \alpha^2 A_{11} ^\text{N} + B_{11} ^\text{N} + \frac{C_{11}^\text{N} }{\alpha} + \frac{D_{11}^\text{N} }{\alpha^3} = 0 \label{eq:3}\\
&V_d^\text{N} \alpha^3 + C_{12}^\text{N}  \alpha^2 + D_{12}^\text{N}  = 0 \label{eq:4}\\
&4 \alpha^5 A_{11}^\text{N}  + \alpha^2 \left(2 \alpha B_{11}^\text{N}  + C_{11} ^\text{N} - C_{12}^\text{N} \right) - D_{11} ^\text{N} + D_{12} ^\text{N} = 0 \label{eq:5}\\
&\alpha^5 A_{11}^\text{N}  + D_{11}^\text{N}  - \lambda D_{12} ^\text{N} = 0 \label{eq:6}\\
&A_{21} ^\text{N} + B_{21}^\text{N}  + C_{21} ^\text{N} + D_{21} ^\text{N} = 0 \label{eq:7}\\
&5 A_{21} ^\text{N} + 3 B_{21} ^\text{N} = 2 \left(\beta + D_{21}^\text{N} \right) \label{eq:8}\\
&\alpha^7 A_{21} ^\text{N} + \alpha^5 B_{21}^\text{N}  + \alpha^2 C_{21}^\text{N} + D_{21}^\text{N}  = 0 \label{eq:9}\\
&\alpha^2 C_{22} ^\text{N} + D_{22}^\text{N}  = 0 \label{eq:10}\\
&5 \alpha^7 A_{21}^\text{N}  + 3 \alpha^5 B_{21} ^\text{N} - 2 D_{21} ^\text{N} + 2 D_{22} ^\text{N} = 0 \label{eq:11}\\
&8 \alpha^7 A_{21} ^\text{N} + 3 \alpha^2 \left(\alpha^3 B_{21}^\text{N}  + C_{21} ^\text{N} - \lambda C_{22}^\text{N} \right) + 8 D_{21} ^\text{N} - 8 \lambda D_{22} ^\text{N} = 0 \label{eq:12}\\
&C_{11} ^\text{N} = 0 \label{eq:13}\\
&C_{12}^\text{N}  = 0 \label{eq:14}
\end{align}

\section{Expressions for the constants in Eqn.~\ref{eqn:first_order_str_simp}}
\label{sec:AppendixB}

\begin{align}
&g_1 = \frac{8 \gamma \left(35 r^{10} A_{11} ^\text{N} A_{21}^\text{N}  + 14 r^7 A_{21}^\text{N} C_{11} ^\text{N} + 9 C_{21} ^\text{N} \left(r^2 C_{11} ^\text{N} + 10 D_{11}^\text{N} \right)\right) m_1}{5 r^8} \label{eq:a1}\\
&g_2 = \frac{24 \gamma \left(49 r^{12} A_{21}^{\text{N}^2} + 23 r^7 A_{21} ^\text{N}C_{21} ^\text{N}+ r^2 \left(24 C_{21}^{\text{N}^2} + 7 C_{11} ^\text{N}\left(r^5 A_{11} ^\text{N}+ r^2 C_{11} ^\text{N}+ 6 D_{11}^\text{N}\right)\right) + 100 C_{21} ^\text{N}D_{21}^\text{N}\right) m_1}{7 r^9} \label{eq:a2}\\
&g_3 = \frac{48 \gamma \left(C_{21} ^\text{N}\left(10 r^5 A_{11}^\text{N} + 16 r^2 C_{11}^\text{N} + 45 D_{11}^\text{N}\right) + A_{21} ^\text{N}\left(11 r^7 C_{11} ^\text{N}+ 35 r^5 D_{11}^\text{N}\right) + 25 C_{11}^\text{N} D_{21}^\text{N}\right) m_1}{5 r^8} \label{eq:a3}\\
&g_4 = \frac{80 \gamma \left(18 C_{21}^\text{N} \left(r^2 C_{21}^\text{N} + 3 D_{21}^\text{N}\right) + A_{21}^\text{N} \left(33 r^7 C_{21} ^\text{N}+ 49 r^5 D_{21}^\text{N}\right)\right) m_1}{7 r^9} \label{eq:a4}\\
&h_1 = \frac{72 C_{22}^\text{N} \left(r^2 C_{12} ^\text{N}+ 10 D_{12}^\text{N}\right) m_2}{5 r^8} \label{eq:b1}\\
&h_2 =\frac{24 \left(r^2 \left(24 C_{22}^{\text{N}^2} + 7 C_{12}^\text{N} \left(r^2 C_{12} ^\text{N}+ 6 D_{12}^\text{N}\right)\right) + 100 C_{22} ^\text{N}D_{22}^\text{N}\right) m_2}{7 r^9} \label{eq:b2}\\
&h_3 =\frac{48 \left(45 C_{22} ^\text{N}D_{12}^\text{N} + C_{12}^\text{N} \left(16 r^2 C_{22} ^\text{N}+ 25 D_{22}^\text{N}\right)\right) m_2}{5 r^8} \label{eq:b3}\\
&h_4 =\frac{1440 C_{22} ^\text{N}\left(r^2 C_{22}^\text{N} + 3 D_{22}^\text{N}\right) m_2}{7 r^9} \label{eq:b4}
\end{align}

\section{Expressions for the coefficients $\mathcal{I}_{lk}^{\text{De}_2}$ in Eqn.~\ref{eqn:first_order_str_sol}}

\label{sec:AppendixC}

\begin{align}
&\mathcal{I}_{11}^{\text{De}_2} = \frac{\gamma \left(r^{10} A_{11} ^\text{N}A_{21}^\text{N} - 14 r^7 A_{21}^\text{N} C_{11} ^\text{N}+ r^2 C_{11}^\text{N} C_{21} ^\text{N}+ C_{21} ^\text{N}D_{11}^\text{N}\right) m_1}{5 r^4} \label{eq:c1}\\
&\mathcal{I}_{21} ^{\text{De}_2}= \frac{\gamma \left(7 r^{12} A_{21}^{\text{N}^2} + 69 r^7 A_{21}^\text{N} C_{21} ^\text{N}+ 3 r^2 \left(4 C_{21}^{\text{N}^2} + 7 C_{11} ^\text{N}\left(r^5 A_{11} ^\text{N}- r^2 C_{11}^\text{N} + D_{11}^\text{N}\right)\right) + 6 C_{21} ^\text{N}D_{21}^\text{N}\right) m_1}{21 r^5} \label{eq:c2}\\
&\mathcal{I}_{31} ^{\text{De}_2}= \frac{\gamma \left(22 r^7 A_{21}^\text{N} C_{11}^\text{N} + 12 r^5 A_{11}^\text{N} C_{21}^\text{N} - 48 r^2 C_{11} ^\text{N}C_{21}^\text{N} + 42 r^5 A_{21}^\text{N} D_{11}^\text{N} + 27 C_{21}^\text{N} D_{11}^\text{N} + 15 C_{11}^\text{N} D_{21}^\text{N}\right) m_1}{15 r^4} \label{eq:c3}\\
&\mathcal{I}_{41} ^{\text{De}_2}= \frac{2 \gamma \left(11 r^7 A_{21}^\text{N} C_{21}^\text{N} - 27 r^2 C_{21}^{\text{N}^2} + 21 r^5 A_{21} ^\text{N}D_{21} ^\text{N}+ 18 C_{21}^\text{N} D_{21}^\text{N}\right) m_1}{21 r^5} \label{eq:c4}\\
&\mathcal{I}_{12}^{\text{De}_2} = \frac{C_{22}^\text{N} \left(r^2 C_{12}^\text{N} + D_{12}^\text{N}\right) m_2}{5 r^4} \label{eq:d1}\\
&\mathcal{I}_{22} ^{\text{De}_2}= \frac{\left(-7 r^4 C_{12}^{\text{N}^2} + 7 r^2 C_{12}^\text{N} D_{12}^\text{N} + 2 C_{22}^\text{N} \left(2 r^2 C_{22}^\text{N} + D_{22}^\text{N}\right)\right) m_2}{7 r^5} \label{eq:d2}\\
&\mathcal{I}_{32} ^{\text{De}_2}= -\frac{\left(16 r^2 C_{12}^\text{N} C_{22}^\text{N} - 9 C_{22}^\text{N} D_{12}^\text{N} - 5 C_{12}^\text{N} D_{22}^\text{N}\right) m_2}{5 r^4} \label{eq:d3}\\
&\mathcal{I}_{42} ^{\text{De}_2}= -\frac{6 C_{22} ^\text{N}\left(3 r^2 C_{22}^\text{N} - 2 D_{22}^\text{N}\right) m_2}{7 r^5} \label{eq:d4}
\end{align}

\newpage
\section{Case 1: $\mathcal{O}(\text{De}_1)$ correction to the pressure field}

\label{sec:AppendixD}

\begin{figure}[!tbh]
  \centering
  \includegraphics[width=\linewidth]{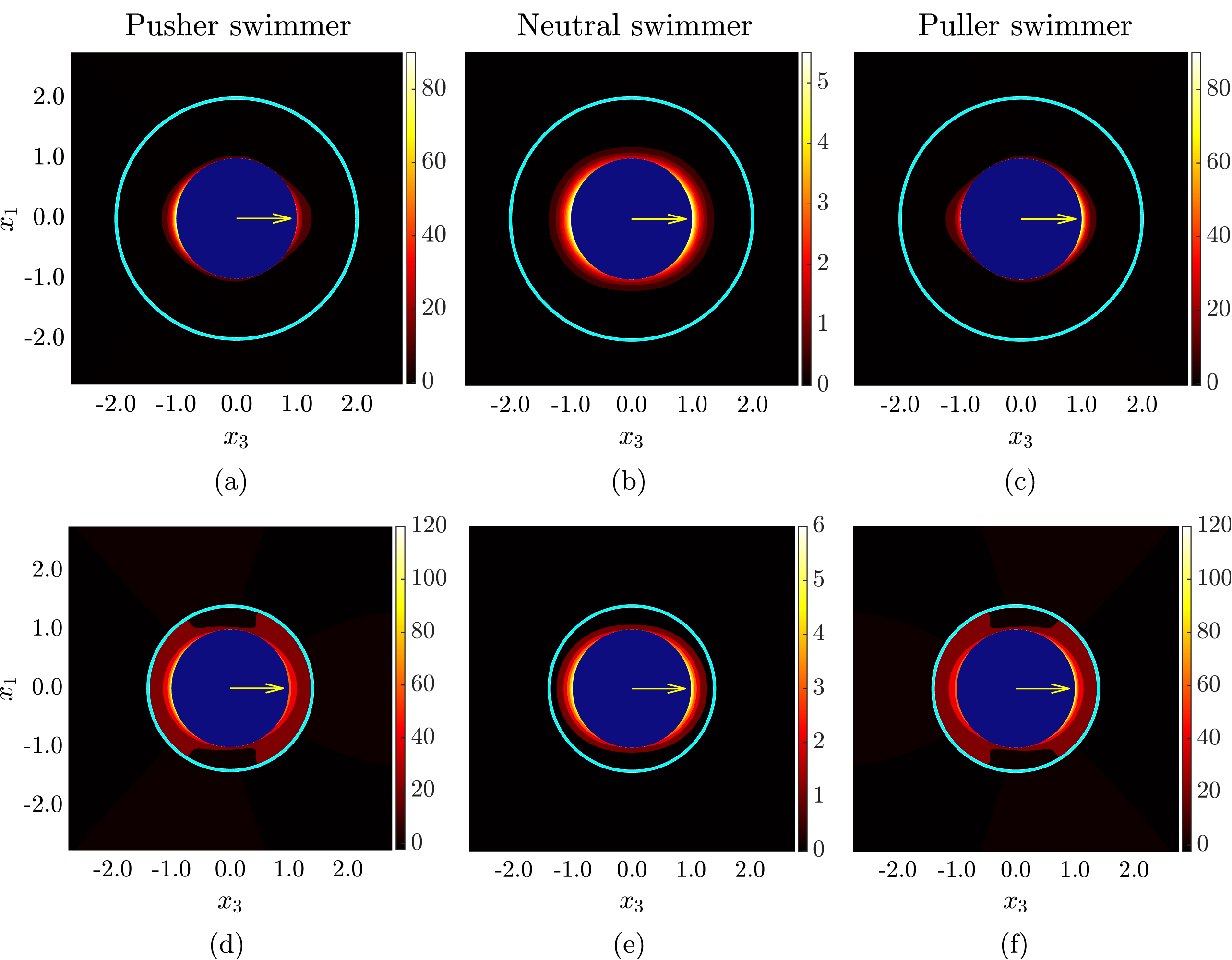}
  \caption{\label{fig:case1_PF} $\mathcal{O}(\text{De}_1)$ correction to the pressure field of an active compound particle when the droplet fluid is viscoelastic and the surrounding fluid is Newtonian. The cyan circle represents the droplet interface, and the black-filled circle indicates the squirmer, with its orientation marked by a yellow arrow. White lines with arrows are streamlines. The top row (a-c) corresponds to a size ratio \(\alpha = 2\), and the bottom row (d-f) corresponds to a size ratio \(\alpha = 1.4\). The first column (a, d) represents pusher swimmers, the middle column (b, e) are neutral swimmers, and the last column (c, f) are puller swimmers.}
\end{figure}

\newpage
\section{Case 2: Forces acting on the squirmer}
\label{sec:AppendixE}

\begin{figure}[!tbh]
  \centering
  \includegraphics[width=0.45\linewidth]{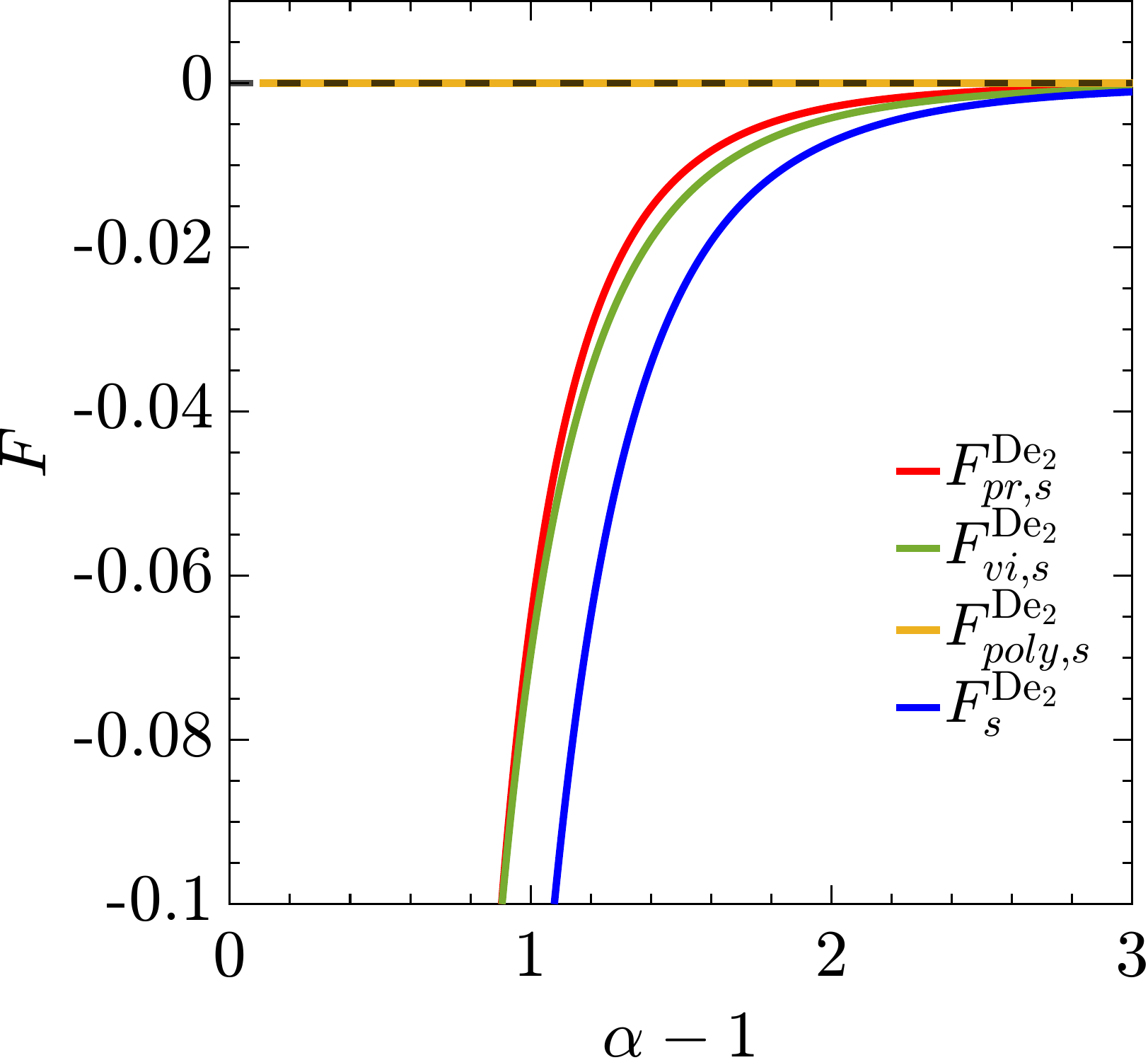}
  \caption{\label{fig:case2_forces} For Case 2, where the surrounding fluid is viscoelastic, the different contributions to the force acting on the squirmer in a fluid pumping problem as a function of \(\alpha - 1\) for \(\lambda = 1\) and \(\beta = -3\). The total force (\(F_{s}^{\text{De}_2}\)) includes contributions from polymeric stress based on the \(\mathcal{O}(1)\) flow field (\(F_{poly,s}^{\text{De}_2}\)), as well as \(\mathcal{O}(\text{De}_2)\) pressure (\(F_{pr,s}^{\text{De}_2}\)) and velocity (\(F_{vi,s}^{\text{De}_2}\)) fields. The dashed black line indicates the zero-force. Here, $m_1 = 0.0$ and $m_2 = 1.0$.}
\end{figure}

\providecommand*{\mcitethebibliography}{\thebibliography}
\csname @ifundefined\endcsname{endmcitethebibliography}
{\let\endmcitethebibliography\endthebibliography}{}

\end{document}